\newcommand\ha{{H$\alpha$}}

\newcommand\kms{\:\rm{km\,s^{-1}}}

\newcommand\LUM{\:{\rm erg\:s^{-1}}}

\newcommand\VEL{\:{\rm km\:s^{-1}}}

\newcommand\SiiL{[\ion{S}{2}] $\lambda\lambda 6716, 6731$}
\newcommand\NiiL{[\ion{N}{2}] $\lambda\lambda 6548, 6583$}
\newcommand\OiLL{[\ion{O}{1}] $\lambda\lambda 6300, 6363$}

\newcommand\sii{[\ion{S}{2}]}
\newcommand\nii{[\ion{N}{2}]}

\newcommand\hii{\ion{H}{2}}

\def\farcm  {\hbox{$.\mkern-4mu^\prime$}}
\def\farcs  {\hbox{$.\mkern-4mu^{\prime\prime}$}}

\documentclass[default]{aastex62}

\received{September 11, 2019}
\revised{October 15, 2019}
\accepted{\today}
\submitjournal{ApJ}

\shorttitle{Separating SNRs from H~II regions in Nearby Galaxies}
\shortauthors{Points et al.}
\usepackage{natbib}
\usepackage{comment}
\usepackage{todonotes}
\usepackage{graphicx}
\begin{document}
%\slugcomment{Draft \today}

\title{Kinematics: A Clean Diagnostic for Separating Supernova Remnants from H~II Regions in Nearby Galaxies}

\author[0000-0002-4596-1337]{Sean D. Points}
\affiliation{National Optical Astronomy Observatory/Cerro Tololo Inter-American Observatory, 
Casilla 603, La Serena, Chile; 
spoints@ctio.noao.edu}

\author[0000-0002-4134-864X]{Knox S. Long}
\altaffiliation{Visiting astronomer, Cerro Tololo Inter-American Observatory, 
National Optical Astronomy Observatory, which is operated by the Association of 
Universities for Research in Astronomy (AURA) under a cooperative agreement
with the National Science Foundation.}
\affil{Space Telescope Science Institute,
3700 San Martin Drive,
Baltimore MD 21218, USA; long@stsci.edu}
\affil{Eureka Scientific, Inc.
2452 Delmer Street, Suite 100,
Oakland, CA 94602-3017}

\author[0000-0001-6311-277X]{P. Frank Winkler}
\altaffiliation{Visiting astronomer, Cerro Tololo Inter-American Observatory, 
National Optical Astronomy Observatory, which is operated by the Association of 
Universities for Research in Astronomy (AURA) under a cooperative agreement
with the National Science Foundation.}
\affiliation{Department of Physics, Middlebury College, Middlebury, VT, 05753; 
winkler@middlebury.edu}

\author[0000-0003-2379-6518]{William P. Blair}
\altaffiliation{Visiting astronomer, Cerro Tololo Inter-American Observatory, 
National Optical Astronomy Observatory, which is operated by the Association of 
Universities for Research in Astronomy (AURA) under a cooperative agreement
with the National Science Foundation.}
\affiliation{The Henry A. Rowland Department of Physics and Astronomy, 
Johns Hopkins University, 3400 N. Charles Street, Baltimore, MD, 21218; 
wblair@jhu.edu}

%\todo[inline]{Finalize Title.}

%% Note that the \and command from previous versions of AASTeX is now
%% depreciated in this version as it is no longer necessary. AASTeX 
%% automatically takes care of all commas and "and"s between authors names.

%% AASTeX 6.1 has the new \collaboration and \nocollaboration commands to
%% provide the collaboration status of a group of authors. These commands 
%% can be used either before or after the list of corresponding authors. The
%% argument for \collaboration is the collaboration identifier. Authors are
%% encouraged to surround collaboration identifiers with ()s. The 
%% \nocollaboration command takes no argument and exists to indicate that
%% the nearby authors are not part of surrounding collaborations.

%% Mark off the abstract in the ``abstract'' environment. 

\begin{abstract}

Many more supernova remnants (SNRs) are now known in external galaxies than in
the Milky Way.  Most of these SNRs have been identified using narrow-band imaging,
separating SNRs from \hii\ regions on the basis of  \sii:H$\alpha$ ratios that
are elevated compared to \hii\ regions.  However, the boundary between SNRs and
\hii\ regions is not always distinct, especially at low surface brightness.
Here we explore  velocity structure as a possible criterion for separating
SNRs from \hii\ regions, using a sample of well-studied SNRs in the Large
Magellanic Cloud (LMC) as well as a small number of SNRs in the galaxy M83\@.
We find, perhaps not surprisingly, that even at large diameters, SNRs exhibit
velocity broadening sufficient to  readily distinguish them from \hii\ regions. 
We thus suggest that the purity of most extragalactic samples would be greatly
improved through spectroscopic observations with a velocity resolution of
order 50~km\,s$^{-1}$. 

\end{abstract}

%% Keywords should appear after the \end{abstract} command. 
%% See the online documentation for the full list of available subject
%% keywords and the rules for their use.

\keywords{galaxies: individual (Large Magellanic Cloud) -- galaxies: ISM  -- supernova remnants}

%% From the front matter, we move on to the body of the paper.
%% Sections are demarcated by \section and \subsection, respectively.
%% Observe the use of the LaTeX \label
%% command after the \subsection to give a symbolic KEY to the
%% subsection for cross-referencing in a \ref command.
%% You can use LaTeX's \ref and \label commands to keep track of
%% cross-references to sections, equations, tables, and figures.
%% That way, if you change the order of any elements, LaTeX will
%% automatically renumber them.

%% We recommend that authors also use the natbib \citep
%% and \citet commands to identify citations.  The citations are
%% tied to the reference list via symbolic KEYs. The KEY corresponds
%% to the KEY in the \bibitem in the reference list below. 

\section{Introduction} \label{sec:intro}

The study of extragalactic supernova remnants (SNRs) began modestly with the
discovery of three radio candidates in the Large  Magellanic Cloud (LMC) by
\citet{mathewson64}, later confirmed through their radio and optical properties
by \citet{westerlund66}.   It was over a decade until SNR studies expanded to
more distant galaxies through the optical identification of  three remnants in
M33 by \citet{dodorico78}.   Great advances have been made since, so that today
there are a total of over 1400 known SNRs (or at least strong candidates) in
about a dozen galaxies, up to distances of $\sim 10$ Mpc---many times the
number known in our own Galaxy \citep[see][for a recent discussion of the state
of play in the field]{long17}.  

Unlike in the Galaxy, where most SNRs were first discovered through their radio
properties and only about a third have even been detected optically
\citep{green17}, the vast majority of extragalactic remnants have been
discovered optically.  Some SNRs have been detected at X-ray and radio wavelengths in 
galaxies outside the Local Group, but due to limited sensitivity and angular resolution  few were originally discovered in these wavelength bands. The classic method to identify SNRs optically is through the \sii:\ha\ ratio,
where shock heated nebulae (primarily SNRs) typically have \sii:\ha\ $\gtrsim
0.4$, compared with photoionized nebulae (\hii\ regions and planetary nebulae)
where \sii:\ha\ is typically $\lesssim 0.2$.   This notable difference results
from the fact that the  gas behind shocks like those characterizing SNRs has a
long cooling tail containing a range of low-ionization species, including
S$^+$, N$^+$, and O$^0$, while the  photoionized gas surrounding hot stars
is generally more highly ionized.  

As larger telescopes and more sensitive instrumentation have enabled
exploration of deeper extragalactic samples, the tidy \sii:\ha\ ratio
criterion for distinguishing SNRs from \hii\ regions has become blurred.   At
low surface brightness especially, photoionized regions such as the warm 
ionized medium in our Galaxy and the so-called diffuse ionized gas (DIG) in 
nearby galaxies 
can display \sii:\ha\ ratios that
approach or exceed 0.4, as density decreases and the distance from the 
ionizing sources increases \citep[e.g.,][]{haffner99}.  Hence, a 
patch of DIG in a nearby galaxy may be misidentified as a potential low-surface-brightness SNR.  Also, extragalactic SNRs may be embedded in
complex emission regions or located so close to regions of \hii\ emission that
they cannot be resolved with ground-based facilities.  Intrinsic 
observational uncertainty in the derived line ratios for faint emission-line 
objects can also cause ambiguity for objects 
with observed ratios near the 0.4 threshold.  Therefore, it would be
valuable to have an additional optical diagnostic for positive identification
of SNRs; hence the motivation for this study.

A potential diagnostic for SNRs is to take advantage of the fact
that the material behind a SNR shock will have bulk motions comparable to the local 
shock speed, several hundred $\rm km~s^{-1}$, whereas material in \hii\ regions 
will for the
most part have velocities of order 10 $\rm km~s^{-1}$.  Unfortunately,
although spectroscopic follow-up of a large number of the nebulae identified as
SNRs in extragalactic surveys has taken place \citep[e.g.,][]{matonick97, blair97, gordon98, lee14a, lee14b, lee15, winkler17, long18,
long19}, most of these spectroscopic
surveys have not had sufficient spectral resolution to measure the expected
line broadening in SNRs and use it as a meaningful diagnostic.

Here we report a survey of selected SNRs in the LMC, along with a few SNR 
candidates
in M83, to test whether measuring line broadening is a practical way to separate
SNRs from \hii\ regions, and to assess what additional information line
broadening may provide in attempting to understand the nature of SNRs in our
sample.  The LMC, with 59 SNRs and 15 additional candidates, is the obvious 
place to carry out such a study because it has
the most fully characterized sample of SNRs of any galaxy \citep[see, e.g.][]{maggi16,bozzetto17}.
There is, however, one obvious observational difference between the SNRs in 
the LMC and those in other more distant galaxies:  at 50 kpc 
\citep{pietrzynski13}, the SNRs in the LMC are fully resolved, and simple 
long-slit spectroscopy samples only a small portion of each SNR.
This is in contrast to  the situation in a galaxy like 
M83 \cite[$d=4.6$ Mpc,][]{saha06}, where a spectroscopic aperture (slit or 
fiber) typically samples most or all of an SNR.  To address this 
situation, we  obtained  spectra of several SNRs and a few \hii\ regions in the 
LMC by scanning the slit over the entire object, thus obtaining an 
integrated spectrum analogous to what a slit might encompass for an object 
in a galaxy like M83. 

In the following sections, we detail our new kinematic observations of SNRs, present our results, and then provide a discussion of our findings.  A brief summary of our conclusions appears at the end.

\section{Observations and Data Reduction}\label{sec:obs}

We used the Goodman High-Throughput Spectrograph \citep[GHTS,][]{clemens04} 
at the SOAR 4.1-m telescope on Cerro Pach\'on to measure the velocity 
broadening of emission lines as a potential diagnostic of SNRs.  
Our observations took place on five nights:  2018 Dec 14-16 and 2019 Jan 15-16
(UT), as detailed in Table~1.  We used the Goodman Red Camera, equipped
with a $4096 \times 4112$ pixel back-illuminated, deep-depletion, astro-multi-2
coated e2v 231-84 CCD, and chose the red-optimized
1200~lines~mm$^{-1}$ Volume Phased Holographic (VPH) grism for our observations.  
At a central wavelength of 6840~\AA, the total wavelength 
coverage was 6230--7460~\AA, which  included lines of 
interest such as \OiLL, \NiiL, \ha, and \SiiL.  We used a $0\farcs6$ slit 
with $1 \times 1$ pixel binning as the optimum compromise between resolution 
and throughput.  The instrumental profile, as measured from the HgAr $+$ Ne
calibration lamp lines was $3.95 \pm 0.17$~pixels ($\approx 1.2$\AA; $\approx 
53\kms$).  The spatial scale along the slit was $0\farcs15$~pixel$^{-1}$.

We obtained spectra
toward a sample of LMC and M83 SNRs and \hii\ regions.  For the primary
LMC targets, we used a drift-scan technique, allowing the narrow slit to 
scan gradually across the entire source.  This enabled us to sample the
full range of filamentary velocities, somewhat analogous to the situation
for long-slit spectra toward M83 and other more distant galaxies, where 
much or all of a marginally resolved source falls within the slit.  We 
selected the LMC targets based on the criteria of size and surface brightness. 
We chose targets that were small enough (in either the N-S and/or E-W 
direction) to be covered in a single scan with the 3\farcm9 long Goodman 
slit, with at least some margin 
at one one or both ends to facilitate sky subtraction.  For simplicity, the slit
was aligned either N-S or E-W so we could scan along a single coordinate 
perpendicular to the slit.  

We used flux-calibrated images 
from the Magellanic Clouds Emission Line Survey \citep[MCELS,][]{smith99, 
winkler15, paredes15} to estimate the total \ha\ flux and surface brightness 
of SNRs in the LMC\@.
Ideally, for this experiment we would have chosen examples covering the full
range from brightest to faintest.  However, because drift-scan spectra are
``inefficient,'' in that much of what the slit sees as it scans over an object
is blank sky rather than filaments of emission, we discovered that it was not
practical to observe the fainter objects with this technique.  

With the Goodman spectrograph, it is quite easy to
switch between imaging and spectrographic modes.  In order to facilitate both
the setups and subsequent analysis, we obtained short direct images in \ha,
SDSS $r^{\prime}$, and in some cases \sii\ as well, before the scanned 
spectra for each object.  

In addition to the scanned spectra, we also obtained long-slit spectra for five
of the LMC SNRs in our sample, in order to assess the effectiveness
of the scans compared with more traditional long-slit spectra.  The positions
and position angles for these long-slit observations were chosen to minimize
stellar contamination while maximizing coverage of the SNR emission regions.
Finally, for comparison with the situation typical for more distant galaxies, 
we also used the same setup to obtain long-slit spectra at two positions in the galaxy M83, chosen to include multiple SNRs from the \citet{blair12}
catalog.  (One or more  \hii\ regions also lay along each slit position.)  
Table 1 gives a journal of all our observations of LMC and M83 targets.  In 
total, we obtained drift-scan spectra toward 15 LMC SNRs and eight LMC \hii\ 
regions (three explicitly targeted and five others covered in SNR scans) plus 
long-slit spectra of five LMC SNRs, five M83 SNRs, and three M83 \hii\
regions.   

After obtaining the spectra for each object, we took comparison lamp spectra
using HgAr $+$ Ne lamps for wavelength calibration.  For flux calibration, we
observed several spectrophotometric standard stars from the list of
\citet{hamuy92, hamuy94} at the beginning and/or end of each night.  

We used common IRAF\footnote{IRAF is distributed by the National Optical
Astronomy Observatory, which is operated by the Association of Universities for
Research in Astronomy (AURA) under a cooperative agreement with the National
Science Foundation.} tasks to perform the basic data reductions, including
bias-subtraction, trimming, and flat-fielding (using dome flats obtained during
the day), for both the drift-scan and long-slit 2-D spectra.

Because the overall exposure time for most of our targets was more than 1800~s,
we investigated the wavelength stability of the data for each object by looking
for differences in the centroids of telluric emission lines.  The positions of
the telluric lines in the object spectrum taken closest in time to the
corresponding comparison lamp spectrum were used as the fiducial values.  If
the telluric emission lines in another spectrum of that object were not well
aligned with the fiducial spectrum, we determined the pixel offset between that
spectrum and the fiducial, shifted the spectrum to match the fiducial, and
combined the individual spectra.  After all the spectra for a given object had
been aligned and combined, we used the comparison spectra and IRAF tasks in the {\tt twodspec.longslit} package to  rectify each 2-D spectrum to a linear wavelength scale.  

We also  
derived a wavelength solution for the all the standard star spectra, combined 
these for each night, and determined a nightly sensitivity function, which we 
applied to all the object spectra for that night. Finally, for each object, we 
then used the IRAF task {\tt background} to subtract the
night-sky emission, selecting as background the portion(s) of the slit that
fell outside the target object.  

%\begin{center}
\begin{deluxetable}{rlllcrrrr}
\tablecaption{SOAR Goodman Observation Log }
\tablehead{\colhead{Object\tablenotemark{a}} & 
 \colhead{Other~Names\tablenotemark{b}} & 
 \colhead{R.A.\tablenotemark{c}} & 
 \colhead{Decl.\tablenotemark{c}} & 
 \colhead{Obs.~Date} & 
 \colhead{Slit PA\tablenotemark{d}} & 
 \colhead{Exposure} & 
 \colhead{Mode\tablenotemark{e}} & 
 \colhead{Scan~Rate} 
\\
\colhead{~} & 
 \colhead{~} & 
 \colhead{(J2000)} & 
 \colhead{(J2000)} & 
 \colhead{(UT)} & 
 \colhead{($\deg$)} & 
 \colhead{(s)} & 
 \colhead{Scan/LS} & 
 \colhead{(\arcsec~hr$^{-1}$)} 
}
\tabletypesize{\scriptsize}
\tablewidth{0pt}\startdata
\bf{LMC SNRs:} \\
N4 &  J0453-6655 &  04:53:14.0 &  -66:55:10 &  2018-12-16 &  0 &  2x1800 &  Scan &  400 \\ 
N11L &  J0454-6625 &  04:54:49.9 &  -66:25:36 &  2018-12-15 &  0 &  3x1800 &  Scan &  122 \\ 
 &   &  04:54:50.7 &  -66:25:50 &  2019-01-15 &  43 &  2x900 &  LS &  0 \\ 
N86 &  J0455-6839   &  04:55:43.7 &  -68:39:02 &  2018-12-14 &  90 &  3x1200 &  Scan &  400 \\ 
 &   &  04:55:42.8 &  -68:38:06 &  2019-01-16 &  74 &  2x1000 &  LS &  0 \\ 
N103B &  J0508-6843 &  05:08:59.4 &  -68:43:35 &  2018-12-14 &  0 &  3x1200 &  Scan &  60 \\ 
N120 &  J0518-6939 &  05:18:43.5 &  -69:39:11 &  2018-12-16 &  0 &  3x1200 &  Scan &  225 \\ 
 &   &  05:18:51.0 &  -69:39:13 &  2018-12-16 &  14 &  2x600 &  LS &  0 \\ 
B0520-69.4 &  J0519-6926 &  05:19:45.3 &  -69:26:01 &  2019-01-15 &  90 &  3x1800 &  Scan &  280 \\ 
 &   &  05:19:46.8 &  -69:25:38 &  2019-01-15 &  100 &  2x1200 &  LS &  0 \\ 
N132D &  J0525-6938 &  05:25:02.7 &  -69:38:33 &  2019-01-16 &  90 &  2x1800 &  Scan &  210 \\ 
N49B &  J0525-6559 &  05:25:24.9 &  -65:59:18 &  2018-12-14 &  0 &  3x1200 &  Scan &  110 \\ 
 &   &  05:25:24.9 &  -65:59:18 &  2019-01-16 &  328 &  2x900 &  LS &  0 \\ 
N49 &  J0526-6605 &  05:26:00.1 &  -66:05:00 &  2018-12-14 &  0 &  3x600 &  Scan &  600 \\ 
DEM-L205 &  J0528-6726 &  05:28:11.1 &  -67:26:49 &  2018-12-16 &  0 &  3x1800 &  Scan &  500 \\ 
N206 &  J0531-7100 &  05:31:57.9 &  -71:00:16 &  2018-12-16 &  0 &  2x1800 &  Scan &  350 \\ 
N63A &  J0535-6602 &  05:35:43.8 &  -66:02:13 &  2018-12-14 &  90 &  3x1200 &  Scan &  135 \\ 
N59B &  J0536-6735 &  05:36:04.2 &  -67:35:11 &  2018-12-15 &  0 &  3x1200 &  Scan &  375 \\ 
DEM-L316B &  J0547-6943 &  05:47:00.0 &  -69:42:50 &  2019-01-16 &  0 &  3x1800 &  Scan &  320 \\ 
DEM-L316A &  J0547-6941 &  05:47:20.9 &  -69:41:27 &  2018-12-15 &  0 &  3x1800 &  Scan &  160 \\
\bf{LMC H~II regions:} \\
N80 &  - &  04:54:12.6 &  -68:21:51 &  2019-01-15 &  90 &  3x1000 &  Scan &  165 \\ 
N38 &  - &  05:20:32.9 &  -66:46:46 &  2019-01-16 &  0 &  2x1200 &  Scan &  210 \\ 
HII(N132D)-E &  - &  05:25:02.7 &  -69:38:33 &  2019-01-16 &  90 &  2x1800 &  Scan &  210 \\ 
HII(N132D)-W &  - &  05:25:02.7 &  -69:38:33 &  2019-01-16 &  90 &  2x1800 &  Scan &  210 \\ 
HII(N49B) &  - &  05:25:24.9 &  -65:59:18 &  2018-12-14 &  0 &  3x1200 &  Scan &  110 \\ 
HII(N63A) &  - &  05:35:43.8 &  -66:02:13 &  2018-12-14 &  90 &  3x1200 &  Scan &  135 \\ 
HII(N59B) &  - &  05:36:04.2 &  -67:35:11 &  2018-12-15 &  0 &  3x1200 &  Scan &  375 \\ 
N175 &  - &  05:40:43.0 &  -70:02:30 &  2018-12-15 &  0 &  2x1200 &  Scan &  450 \\ 
\bf{M83 Objects:} \\
B12-023 &  B04-07,S03-007 &  13:36:48.3 &  -29:51:45 &  2019-01-15 &  96 &  2x1800 &  LS &  0 \\ 
B12-039 &  B04-11 &  13:36:50.3 &  -29:52:47 &  2019-01-15 &  96 &  2x1800 &  LS &  0 \\ 
B12-180 &  B04-58,S03-109 &  13:37:07.5 &  -29:51:33 &  2019-01-16 &  96 &  3x1500 &  LS &  0 \\ 
B12-191 &  B04-53,S03-105 &  13:37:08.6 &  -29:51:35 &  2019-01-16 &  96 &  3x1500 &  LS &  0 \\ 
B12-208 &  B04-41,S03-081 &  13:37:11.7 &  -29:51:39 &  2019-01-16 &  96 &  3x1500 &  LS &  0 \\ 
M83-HII-1 &  - &  13:36:48.0 &  -29:52:44 &  2019-01-15 &  96 &  2x1800 &  LS &  0 \\ 
M83-HII-2 &  D83-13 &  13:36:52.9 &  -29:52:51 &  2019-01-15 &  96 &  2x1800 &  LS &  0 \\ 
M83-HII-3 &  M06-34 &  13:37:01.4 &  -29:51:26 &  2019-01-16 &  96 &  3x1500 &  LS &  0 \\ 
\enddata 
\tablenotetext{a}{N: \citet{Henize56} identifications; DEM: \citet{Davies76}; B12: \citet{blair12}; HII: H~II emission sections extracted from locations along the long slits.}
\tablenotetext{b}{J\#: as listed by \citet{bozzetto17}; B04: \citet{blair04}; S03: \citet{Soria03} X-ray source number; D83: \citet{deVaucouleurs83}; M06: \citet{Maddox06} radio source.}
\tablenotetext{c}{Object center position adopted from that of \citet{bozzetto17} for scans; slit center position for long slits; object coordinates for M83 objects.}
\tablenotetext{d}{Astronomical position angle (east from north) of long dimension of slit.}
\tablenotetext{e}{Scan = scanned spectrum; LS = Long-slit.}
\label{table_obs}
\end{deluxetable}
%\end{center}

%{\todo[inline]{I would like to (eventually) see these grouped: LMC SNRs; LMC HII regions; M83 objects.
%We should also LSt forget to add LStes about coords (see the commented-out version) and also one listing refs for Other Names. Finally, this would be cleaner if we had a single entry in columns 1\&2 for objects with both scan and LS spectra.  I have done this in the past, but forget how. -- PFW}}

\section{Results and Analysis}\label{sec:redx}

\begin{figure}
\epsscale{0.9}
\plotone{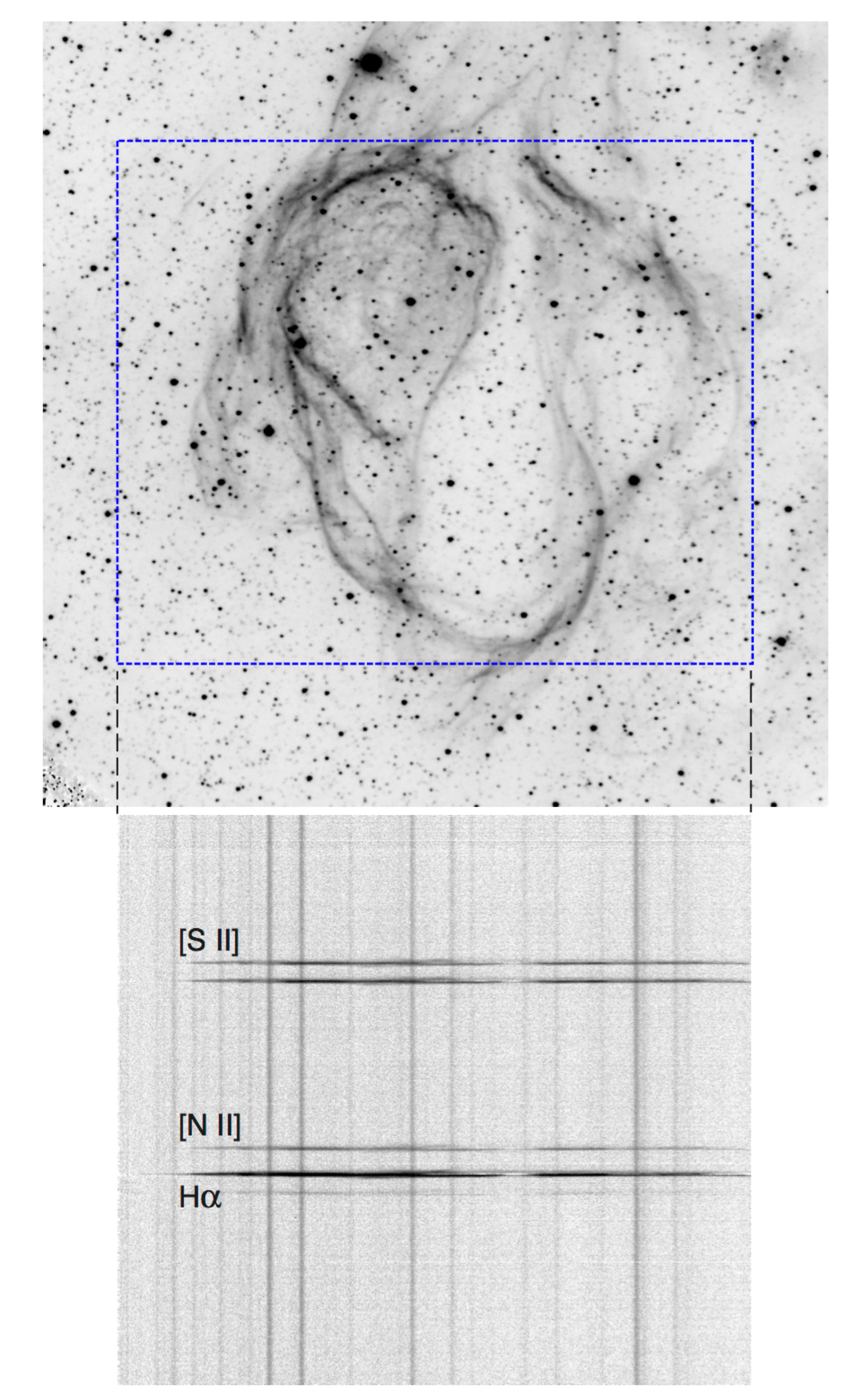}
\caption{({\em top}) An \ha\ image of the LMC SNR N86 (J0455--6839) from the Goodman
spectrograph in direct imaging mode.  The field is 5\arcmin\ square and is
oriented N up, E left.  The blue dashed rectangle indicates the region scanned
north to south using an E-W slit.  ({\em bottom}) A portion of the  sky-subtracted
2-D spectrum from the scan, displayed at the same spatial scale; black dashed
lines in the upper figure indicate the ends of the slit.  Complex velocity
structure, showing both blue- and red-shifts, is present in the nebular
emission.  The dark vertical traces are the signatures of brighter stars within
the scan region, as seen in the image, directly above the spectroscopic traces.
The 1-D extraction for N86 is shown as the second trace from the top in 
Fig.~\ref{fig:example_spectra}.
}
\label{fig:n86_image}
\end{figure}

\begin{figure}
\epsscale{1.1}

\plotone{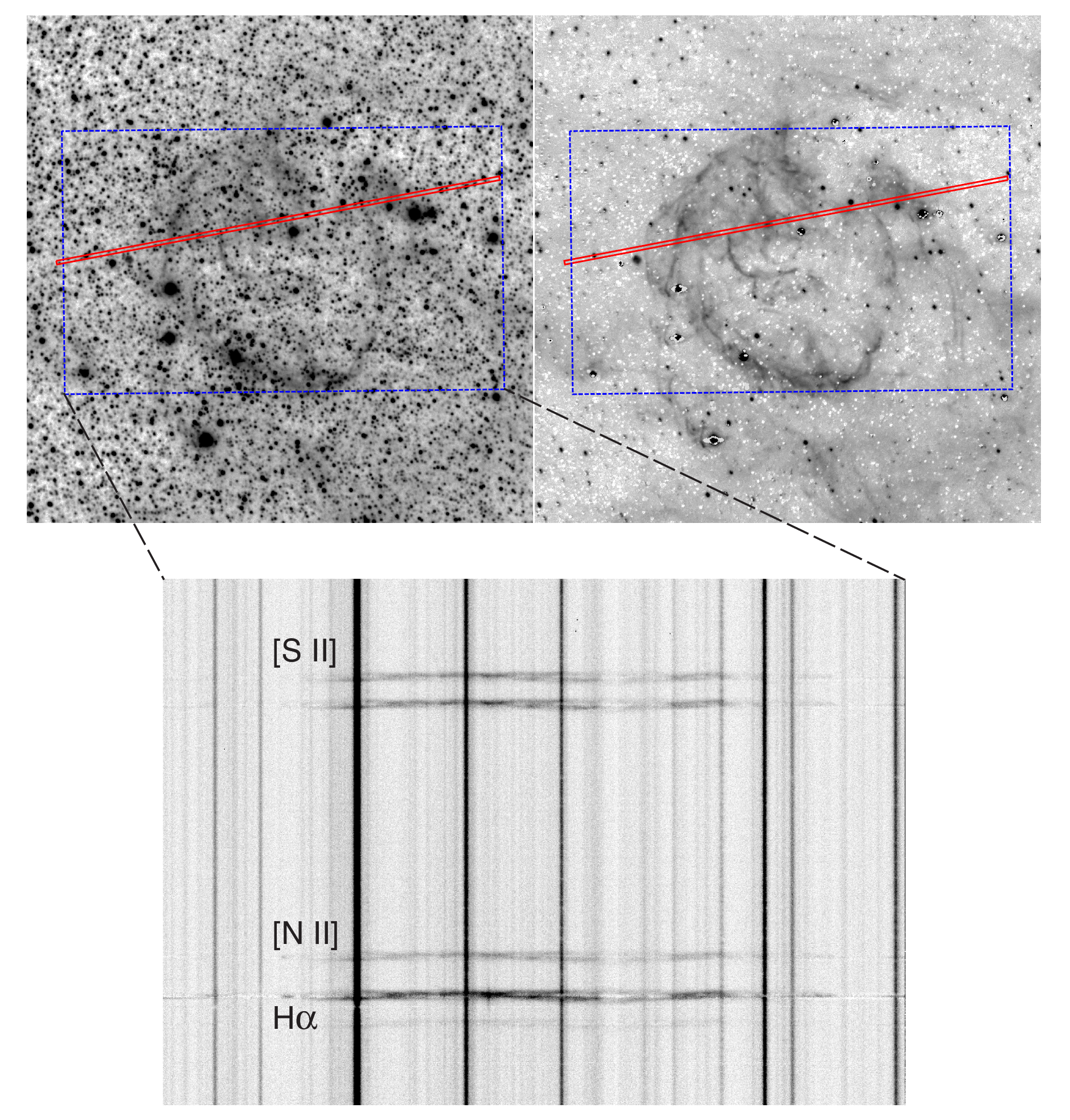}
\caption{({\em top panels}) \ha\ images of the SNR B0520--69.4 = J0519--6926, located in the
densely populated bar of the LMC. The left image is before continuum
subtraction, and the one to the right is after. The field is 4\farcm5 square,
oriented N up, E left.  The blue dashed rectangle indicates the region scanned
north-to-south using an E-W slit.  For this object we also obtained a long-slit spectrum at the position indicated in red.
({\em bottom})  A portion of the 2-D spectrum from the scan over the region indicated above.  Complex velocity
structure along the slit is obvious.  The 1-D  spectrum extracted from this 2-D one is shown as the lower trace of Fig.~\ref{fig:b0520_1d}.
\label{fig:b0520_ha}
}
\end{figure}

Figures~\ref{fig:n86_image} and \ref{fig:b0520_ha} provide representative 
examples of the spectroscopic and imaging data we obtained toward the LMC 
SNR sample.  The upper panel of Figure~\ref{fig:n86_image} shows the \ha\ 
image of the SNR N86 obtained by the Goodman using its direct 
imaging mode, where the blue dashed rectangle outlines the area scanned in 
spectroscopic mode.  Figure~\ref{fig:b0520_ha} shows the \ha\ 
image of the LMC SNR B0520$-$69.4, before (left) and after (right) 
continuum subtraction.
One has no difficulty identifying N86 because of its relatively high 
surface brightness filaments and its location in a sparse field on the 
western edge of the LMC\@.  However, the crowded star field and the relatively 
low surface brightness of the emitting 
filaments in SNR B0520$-$69.4 (see Fig.\ref{fig:b0520_ha} top) make it more 
difficult to identify its structure until after continuum subtraction.  

The background-subtracted 2-D spectra of 
N86 and B0520$-$69.4 (lower panels of Figures~\ref{fig:n86_image} and
\ref{fig:b0520_ha}, respectively) reveal complex kinematic structure, 
despite the crowded
star field in the latter case.   Both of these show multiple red- and
blue-shifted components, especially near the slit centers that tracked across
the central part of the objects, where we expect the line-of-sight velocities
to be highest.  Such complex structure is seen in  virtually all of our scanned
data for LMC SNRs. 

While the 2-D spectra are necessary to reveal the full complexity of the
velocity structure, it is also useful to extract 1-D spectra from the scanned
data in order to provide a single quantitative measure for the velocity
broadening.  The 1-D spectra from LMC objects also provide a direct comparison
with spectra from SNRs and other objects in M83 
and other more distant galaxies.  Therefore, for each of the scanned spectra, 
we summed over all the lines containing the object, omitting small regions 
where bright stars crossed the slit.  
In four of the spectra of our LMC primary targets (N49B, N59B, N63A, 
and N132D) we also detected a second component attributable to 
nearby diffuse \ha\ emission.  In these cases we extracted 1-D spectra 
for both the SNR and \hii\ region components separately.

\begin{figure}
%\epsscale{0.7}
\plotone{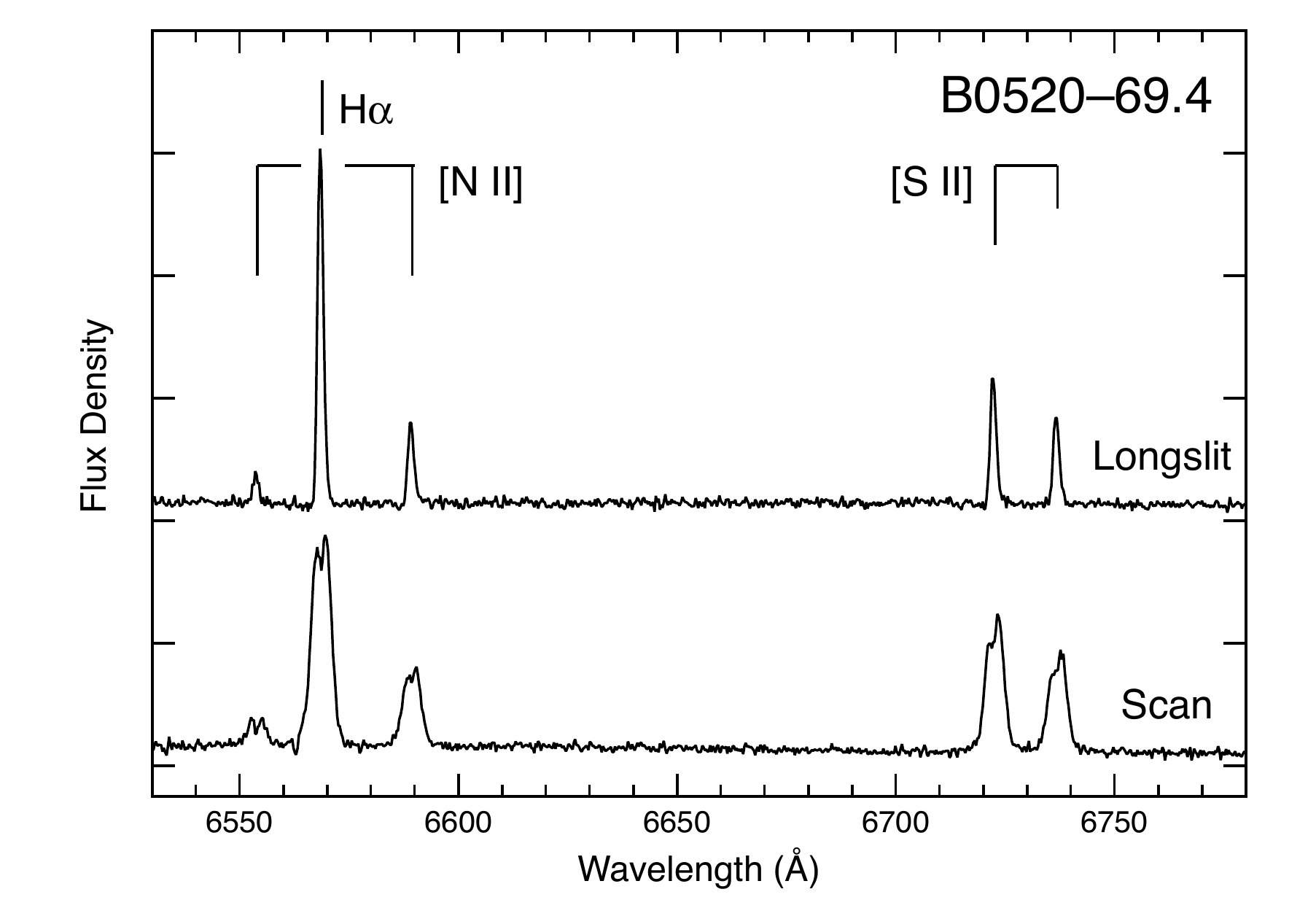}
\caption{The lower trace shows the 1-D spectrum of the SNR B0520--69.4 = J0519--6926 obtained by summing
over the spatial direction in 2-D spectrum from a scan of the entire object
(Fig.~\ref{fig:b0520_ha}, bottom panel), while the upper trace shows a similar extraction from a long-slit 2-D spectrum at the position indicated on the upper panels of Fig.~\ref{fig:b0520_ha}.  The scan clearly shows more
velocity structure and broader lines than the long-slit data.
\label{fig:b0520_1d}
}
\end{figure}

Figure~\ref{fig:b0520_1d} shows examples of  extracted 1-D spectra from
B0520$-$69.4.  The lower trace shows the extracted spectrum from the scanned
observation (lower panel of Fig.~\ref{fig:b0520_ha}), while the upper trace is
from the position of the slit indicated on the upper panels of
Fig~\ref{fig:b0520_ha}.   The scanned data clearly reveal a more complex
velocity structure and broader emission lines with multiple peaks than does the
long-slit spectrum.  The same is true for the other objects observed in both
scanned and long-slit modes, as shown in Table~2.  

\begin{figure}
\plotone{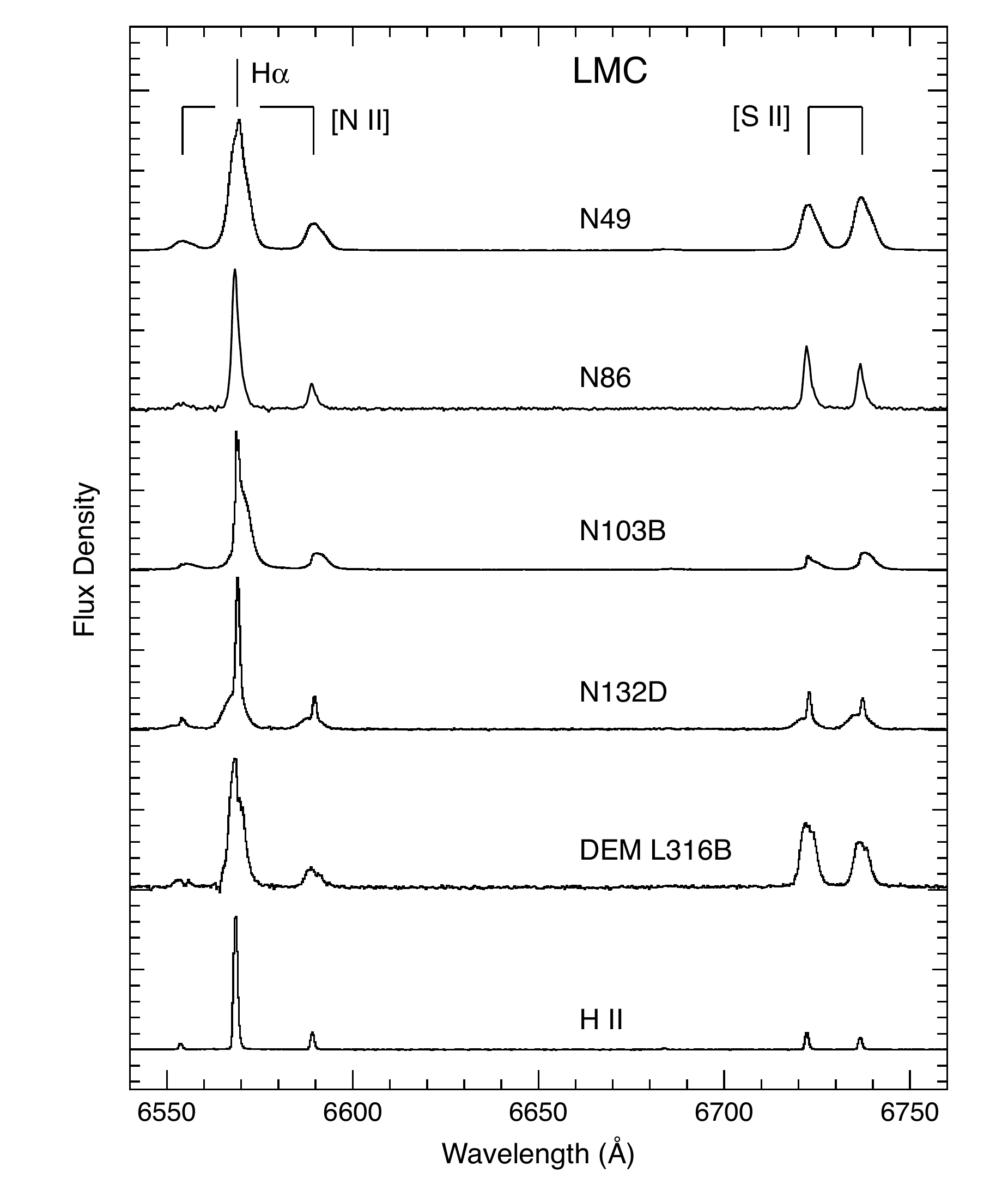}
\caption{Sections from the 1-D spectra for five SNRs and one \hii\ region, all
in the LMC (arbitrarily scaled).  All are the result of scanned 2-D spectra that have been
sky-subtracted, and then summed over the extent of the object to give the 1-D
spectra shown here. All of the SNRs have lines that are noticeably broader then
those in the \hii\ region shown, as well as in other \hii\ regions that we studied. Note that for N103B, the global \sii:\ha\ ratio is less than the canonical value of 0.4 for SNRs, and also that there is a narrow component to the \ha\ line.  Both are due to overlying/contaminating \hii\ emission.  
\label{fig:example_spectra}}
\end{figure}

\begin{figure}
\plotone{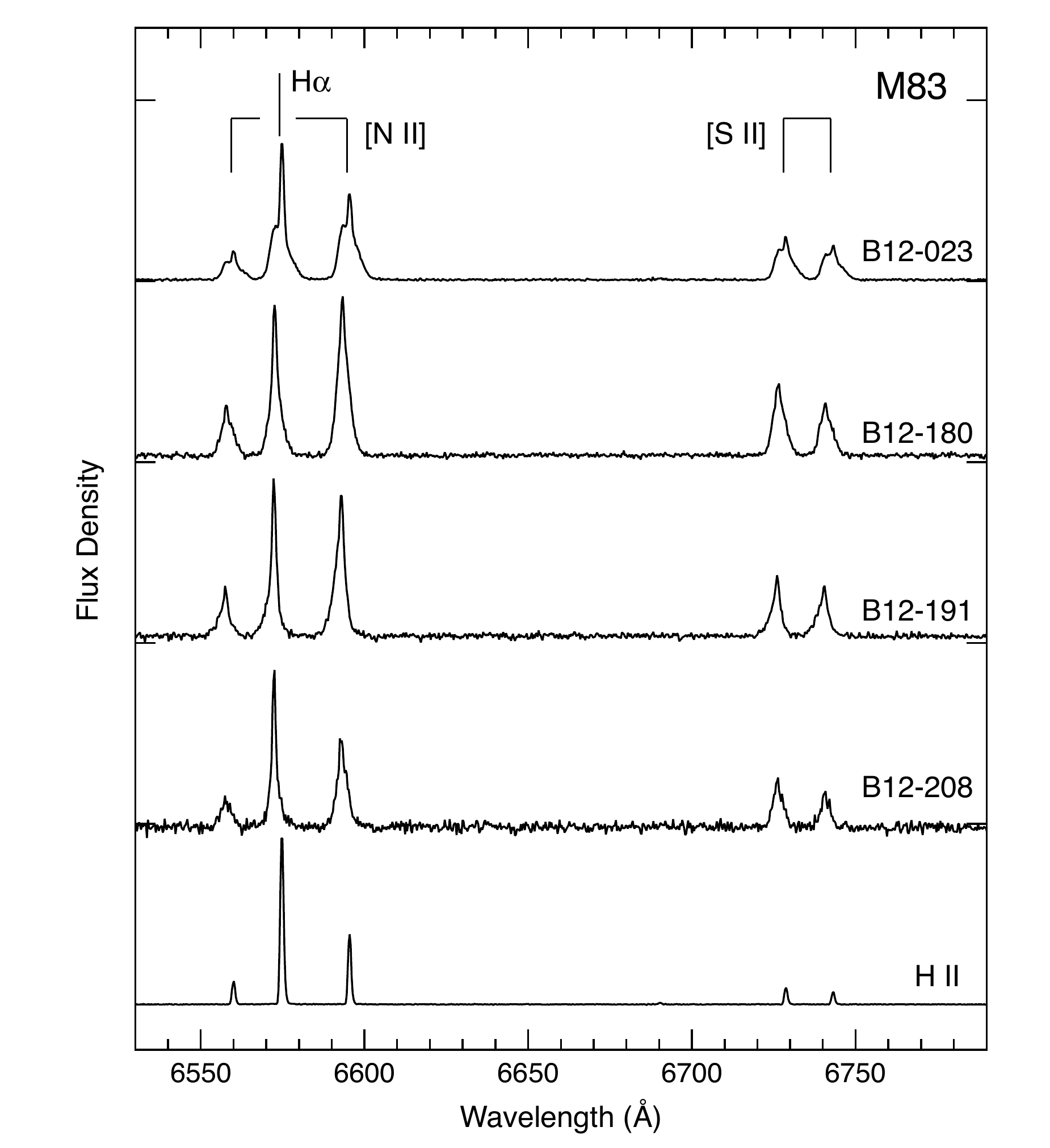}
\caption{Sections from the long-slit spectra for four SNRs in M83, plus an \hii\
region for comparison.  All the SNRs have \sii:\ha\ ratios $>0.4$, as expected,
and all show significantly broadened lines as well.  The lines from the \hii\
region are consistent with the $\sim 53 \kms$ instrumental resolution of the
Goodman spectrograph in the setup we used. 
\label{fig:m83_spectra}}
\end{figure}

For the two slit positions we observed in M83, each object crossed by the slit
(SNRs and serendipitous \hii\ regions) extended over only a few lines
($\lesssim 2\arcsec$) along the slit.  Here we simply summed over the relevant
lines to obtain the 1-D spectra.  Several examples of the extracted spectra
from LMC and M83 SNRs and 
\hii\ regions are shown in Figures~\ref{fig:example_spectra} and  
\ref{fig:m83_spectra}, respectively.  The extracted 1-D  spectra from the 
scanned LMC SNR sample and from the M83 long-slit SNR sample both show the 
same general pattern of emission lines that are broadened in comparison
to those in the
spectra of \hii\ regions. 

To obtain a quantitative comparison between SNRs and \hii\ regions, we 
determined the spectral widths of the emission lines in our sample using two 
different methods for both the scanned and long-slit data.  In the first 
method we simply measured the widths of the lines at 20\% of the peak 
value, while for the second we fit each of the principal lines in the spectra---\ha, 
the \nii\ doublet, and the \sii\ doublet---with simple, single-component 
Gaussians.   Simply measuring the width of the 
profile at a particular percentage of the peak has the advantage that it  
is relatively insensitive to asymmetries in the profiles, but it is quantized 
at the scale of an individual pixel.  Gaussian fits do not suffer from this 
limitation;  they readily give measures for the total 
flux in the lines; and they make comparisons with spectra obtained from 
different instruments more straightforward.   In Table~2, we present the 
20\% results only 
for \ha, while we tabulate the FWHM results for all three lines, both for SNRs 
and \hii\ regions.  As expected, the \ha\ 20\% widths are larger than the 
FWHM values for individual objects, and the 20\% widths are significantly 
higher for SNRs than for \hii\ regions. 

%\begin{center}
\begin{deluxetable}{rrrccrcrcrr}
\tablecaption{Extracted Fluxes and Line Widths }
\tablehead{\colhead{Object} & 
 \colhead{Mode} & 
 \colhead{H$\alpha$} & 
 \colhead{H$\alpha$} & 
 \colhead{H$\alpha$} & 
 \colhead{\nii} & 
 \colhead{\nii} & 
 \colhead{\sii} & 
 \colhead{\sii} & 
 \colhead{L$_{\rm H\alpha}$$^d$} 
\\
\colhead{~} & 
 \colhead{Scan/LS} &
 \colhead{Flux$^a$} & 
 \colhead{(FWHM)$^b$} & 
 \colhead{(20\% Width)$^c$} & 
 \colhead{Flux$^a$} & 
 \colhead{(FWHM)$^b$} &
 \colhead{Flux$^a$} &
 \colhead{(FWHM)$^b$} & 
 \colhead{~} 
 \\
\colhead{~} & 
 \colhead{~} & 
 \colhead{~} & 
 \colhead{($\rm\AA$)} & 
 \colhead{($\rm\AA$)} & 
 \colhead{~} & 
 \colhead{($\rm\AA$)} & 
 \colhead{~} & 
 \colhead{($\rm\AA$)~} & 
 \colhead{~} 
}
\tabletypesize{\scriptsize}
\tablewidth{0pt}\startdata
\bf{LMC SNRs:} \\
N4 &  Scan &  23.00 &  2.35 &  3.62 &  4.03 &  2.39 &  13.90 &  2.39 &  22.8 \\ 
N11L &  Scan &  35.60 &  2.83 &  4.53 &  7.09 &  2.89 &  26.90 &  2.78 &  10.8 \\ 
 &  LS &  90.50 &  2.59 &  3.63 &  17.10 &  2.63 &  66.80 &  2.50 &  -- \\ 
N86 &  Scan &  62.00 &  2.40 &  3.93 &  11.20 &  2.46 &  45.40 &  2.30 &  41.1 \\
 &  LS &  108.00 &  1.87 &  3.02 &  16.70 &  1.80 &  68.50 &  1.87 &  -- \\ 
N103B &  Scan &  43.30 &  4.03 &  5.44 &  7.57 &  4.53 &  11.99 &  3.90 &  4.3 \\ 
N120 &  Scan &  190.00 &  1.67 &  2.72 &  36.30 &  1.87 &  96.80 &  2.75 &  70.8 \\ 
 &  LS &  114.00 &  1.94 &  3.02 &  29.00 &  1.94 &  107.50 &  2.04 &  -- \\ 
B0520-69.4 &  Scan &  18.70 &  4.94 &  6.04 &  6.24 &  4.65 &  17.78 &  4.54 &  13.0 \\ 
 &  LS &  9.58 &  1.55 &  2.42 &  2.25 &  1.59 &  6.05 &  1.58 &  -- \\ 
N132D$^e$ &  Scan &  28.10 &  1.56 &  2.42&  5.94 & 1.58 &  15.50 &  1.40 &  14.8 \\ 
N49B &  Scan & 27.90 &  5.58 &  5.44 &  9.82 &  3.81 &  14.30 &  4.61 &  5.1 \\ 
 &  LS &  31.70 &  5.83 &  5.43 &  16.00 &  4.86 &  19.90 &  4.78 &  -- \\ 
N49 &  Scan &  468.00 &  5.28 &  7.22 &  106.00 &  5.52 &  355.00 &  5.03 &  232.0 \\ 
DEML-205 &  Scan &  189.00 &  1.66 &  2.72 &  27.30 &  1.64 &  71.40 &  1.82 &  235.0 \\ 
N206 &  Scan &  55.30 &  1.63 &  2.42 &  12.10 &  1.85 &  44.00 &  2.56 &  48.1 \\ 
N63A &  Scan &  149.00 &  2.21 &  3.93 &  26.70 &  2.34 &  128.80 &  1.88 &  33.3 \\ 
N59B &  Scan &  67.30 &  1.88 &  2.72 &  10.20 &  1.81 &  32.40 &  1.87 &  41.8 \\ 
DEM-L316B &  Scan &  27.70 &  4.28 &  5.76 &  4.54 &  4.72 &  27.50 &  4.50 &  22.0 \\ 
DEM-L316A &  Scan &  32.70 &  3.73 &  3.93 &  5.54 &  4.07 &  27.50 &  3.48 &  13.0 \\ 
\bf{LMC H~II Regions:} \\
N80 &  Scan &  74.30 &  1.11 &  1.81 &  9.49 &  1.24 &  15.18 &  1.28 &  16.9 \\ 
N38 &  Scan &  297.00 &  1.27 &  1.81 &  22.70 &  1.18 &  33.50 &  1.17 &  103.0 \\ 
HII(N132D)-E &  Scan &  3.09 &  1.34 &  2.11 &  0.54 &  1.19 &  1.17 &  1.21 &  1.6 \\ 
HII(N132D)-W &  Scan &  2.56 &  1.30 &  2.11 &  0.62 &  1.16 &  1.32 &  1.16 &  1.3 \\ 
HII(N49B) &  Scan &  26.90 &  1.18 &  1.81 &  6.27 &  1.18 &  17.11 &  1.12 &  4.9 \\ 
HII(N63A) &  Scan &  70.80 &  1.20 &  1.81 &  7.94 &  1.11 &  10.84 &  1.15 &  15.8 \\ 
HII(N59B) &  Scan &  351.00 &  1.53 &  2.11 &  33.50 &  1.52 &  78.70 &  1.58 &  218.0 \\ 
N175 &  Scan &  130.00 &  1.25 &  1.81 &  15.30 &  1.15 &  26.70 &  1.16 &  96.9 \\ 
\bf{M83 Objects:} \\
B12-023 &  LS &  1.79 &  3.66 &  5.43 &  1.89 &  5.40 &  1.78 &  5.54 &  -- \\ 
B12-039 &  LS &  4.06 &  1.59 &  2.41 &  2.74 &  1.83 &  2.45 &  1.94 &  -- \\ 
B12-180 &  LS &  0.77 &  2.87 &  4.83 &  1.16 &  3.99 &  0.97 &  4.05 &  -- \\ 
B12-191 &  LS &  0.48 &  2.14 &  4.22 &  0.70 &  3.52 &  0.50 &  2.89 &  -- \\ 
B12-208 &  LS &  0.27 &  1.67 &  2.71 &  0.29 &  3.52 &  0.26 &  3.39 &  -- \\ 
M83-HII-1 &  LS &  25.90 &  1.53 &  2.41 &  7.96 &  1.39 &  4.64 &  1.46 &  -- \\ 
M83-HII-2 &  LS &  3.67 &  1.19 &  1.81 &  1.42 &  1.10 &  0.58 &  1.11 &  -- \\ 
M83-HII-3 &  LS &  0.52 &  1.35 &  2.11 &  0.22 &  1.45 &  0.12 &  1.32 &  -- \\ 
\enddata 
\tablenotetext{a}{Fluxes are in units of 10$^{-15}$ ergs cm$^{-2}$ s$^{-1}$. \nii\ flux is for $\lambda$6583 only while \sii\ flux is total for the doublet.  Note that these do {\em not} represent the total flux from the entire object, but for scans is an average flux through the slit as it moved across the object.  For long-slit spectra, these are simply the total flux through the fixed 0\farcs 6 slit.  See text, section 4.2.  }
\tablenotetext{b}{Single Gaussian FWHM values in \AA; \nii\ is for $\lambda$6583 and \sii\ is for $\lambda$6716; see text.}
\tablenotetext{c}{Line widths at 20\% of peak value, in \AA; see text.}
\tablenotetext{d}{The H$\alpha$ luminosities are in units of 10$^{35}$ ergs s$^{-1}$.}
\tablenotetext{e}{N132D is embedded in a region of strong \hii\ emission, which gives a strong narrow component to the lines, producing artificially narrow Gaussian fits.  Fainter broad emission is evident in both the 2-D and 1-D spectra (see Fig.~4).}
\label{table_spec}
\end{deluxetable}
%\end{center}

\begin{figure}
\plottwo{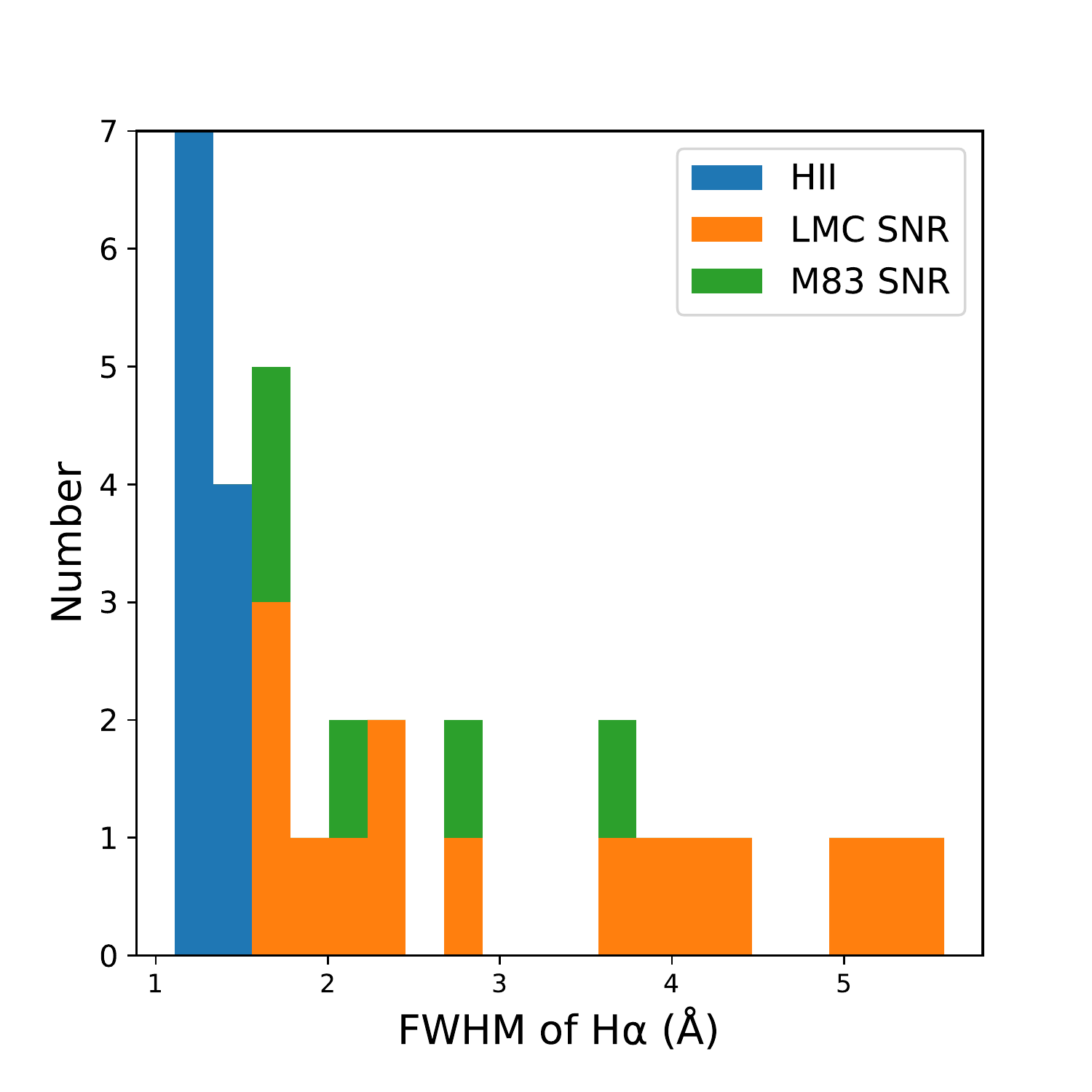}{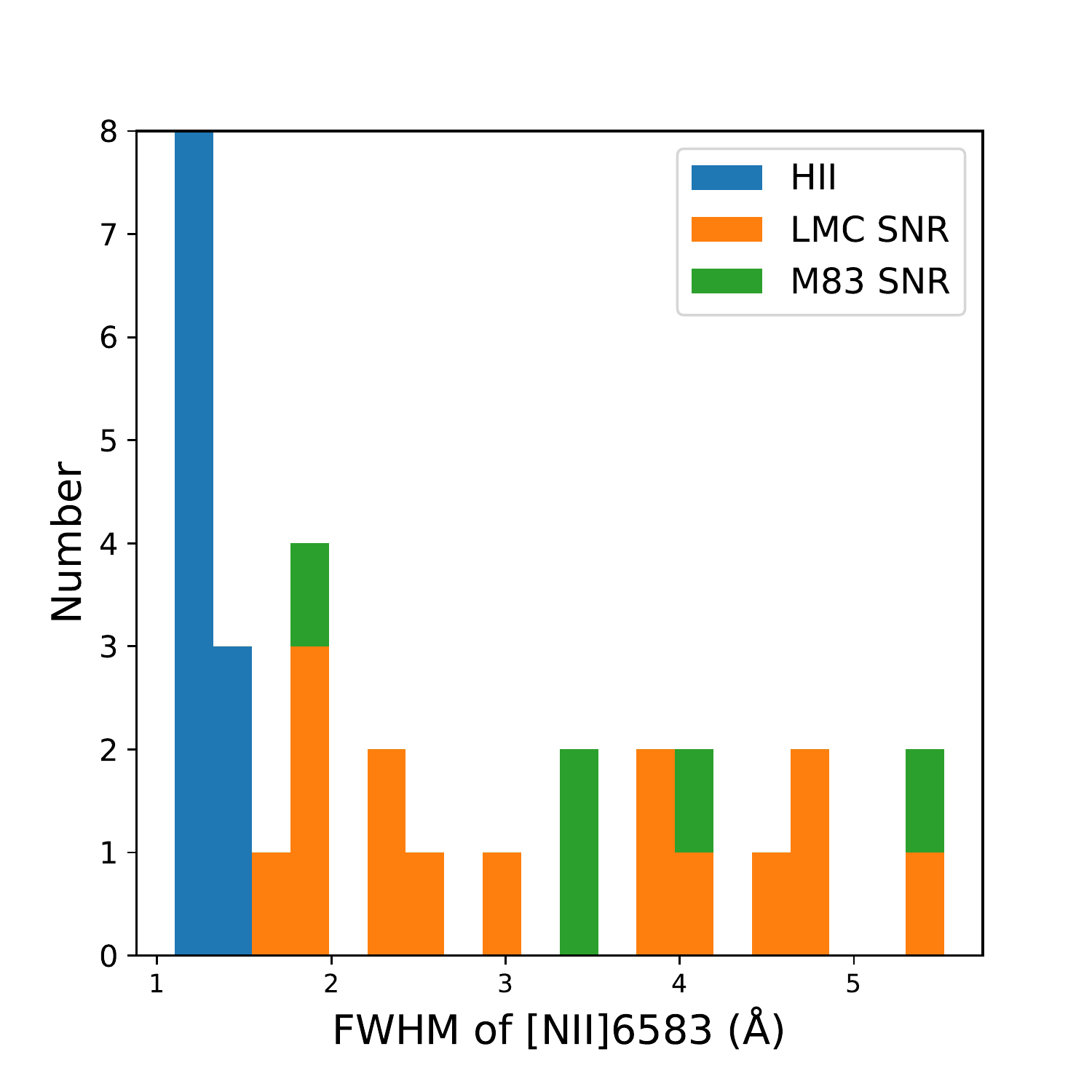}
\caption{ Histograms of the widths of the \ha\ (left) and \nii\  $\lambda$6583 lines in \hii\ region spectra compared to SNR
spectra in the LMC and M83.  Even though the \hii\ region lines are not fully
resolved, the larger values for the SNRs is clear.}
\label{fig:histogram}
\end{figure}

\begin{figure}
\plottwo{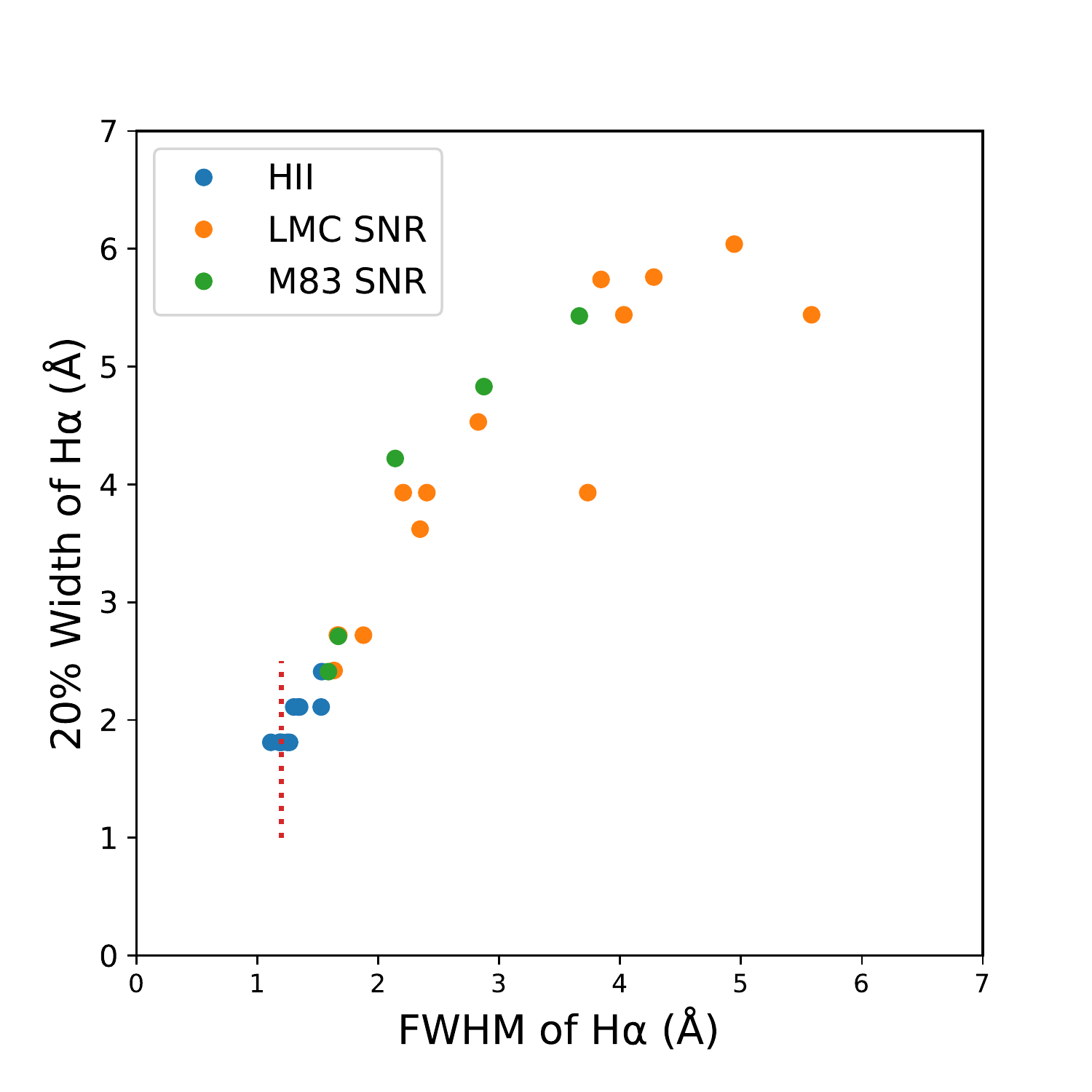}{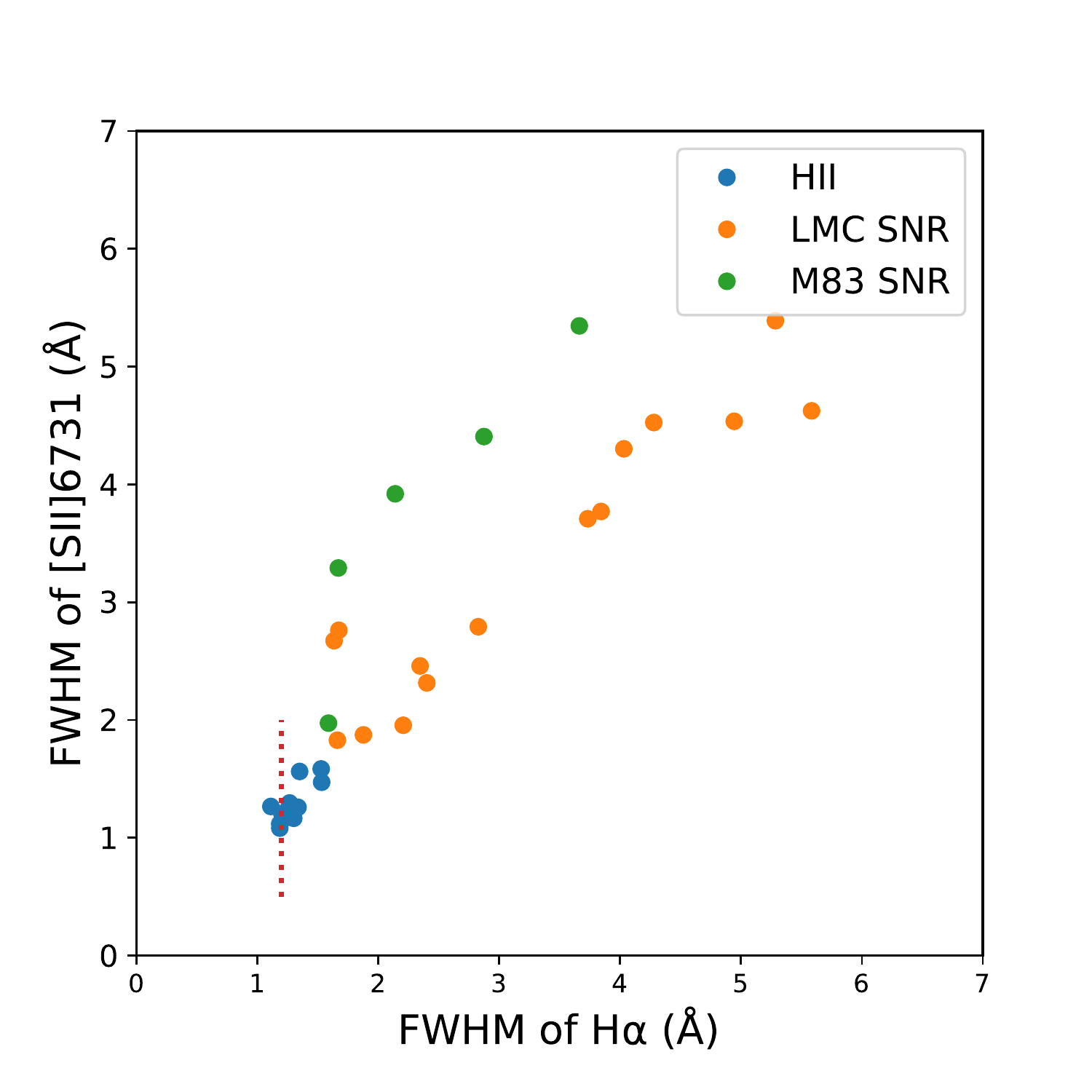}
\caption{Left: A comparison between the width of the lines measured by Gaussian
fitting and the width measured at the 20\% of peak level.  Right: A comparison
between the widths measured by Gaussian fits in \ha\ compared to one of the
[S~II] lines.  For the LMC SNRs, only the scanned spectra are shown. The dotted line at 1.2\AA\ indicates the FWHM of the instrumental resolution. Note the systematically broader [S~II] lines compared with \ha\ for the M83 objects in the right panel.
\label{fig:widths}}
\end{figure}

From the data presented in Table~2, we conclude the following:
%Table~\ref{table_spec}

\begin{itemize}
\item In general the \hii\ regions have narrower line profiles
% measured by Gaussians 
than do the SNRs.  This can be seen qualitatively in 
Figs~\ref{fig:example_spectra} \& \ref{fig:m83_spectra} where the 
emission lines in the SNRs are broader than the emission lines in the \hii\ 
regions. In Figure~\ref{fig:histogram}, we plot the spread of values for the
FWHM from the \ha\ line for the scanned LMC SNRs and \hii\ regions and the
long-slit data of M83 SNRs and \hii\ regions.  As seen in this figure, the 
\hii\ region line profiles are close to the instrumental resolution of ~1.2 
\AA.  Therefore, with sufficiently high spectral resolution, it is easy to 
separate the \hii\ regions from the SNRs (see \S4.1 for details).

\item The FWHM and 20\% widths of the \ha\ line give similar results and are
correlated, as seen in Table~2 and in the left-hand side of 
Figure~\ref{fig:widths}.  In general the FWHM of the \ha\ line is smaller than 
the 20\% width of the \ha\ line.  Hence, even though the Gaussian fits to
the line profiles do not capture the full complexity, and in some case extent, 
of the line profiles, they can be used effectively to discriminate between \hii\ regions 
and SNRs.

\item Our experiments with long-slit spectra make it clear that the kinematic
distinction between SNRs and \hii\ regions is less effective when
observing single positions of spatially-resolved SNRs in long-slit mode than 
it is when observing the entire SNR (cf. Fig.~\ref{fig:b0520_1d}).  
This is probably what one would have expected {\em a priori}, since one generally 
selects slit positions to overlay the brightest portion of a SNR, often along the 
rim with high surface brightness but relatively low line-of-sight shock 
velocity.

\item For the LMC SNRs, the FWHM of \ha\ and other lines are comparable 
(see Fig~\ref{fig:widths}); however, in M83 there is a clear tendency for the 
\sii\ and \nii\ line widths to be greater than those of \ha, indicative 
of contamination of the \ha\ line by adjacent or overlying photoionized 
emission that could not be accurately subtracted.  This indicates that 
velocity broadening of forbidden lines may be the preferred diagnostic in 
galaxies beyond the Local Group.

\end{itemize}

\section{Discussion}

We have obtained drift-scan spectra toward 15 SNRs and eight \hii\ regions in 
the LMC in addition to long-slit spectra of five SNRs and three \hii\ regions
in M83\@.  We scanned across the LMC targets in order to measure the integrated 
line profiles of the SNRs and \hii\ regions so that we can compare the results
to the measured properties of SNRs and \hii\ regions in more distant galaxies
where objects can be centered in a single long-slit spectrum.  Below we 
discuss our results in detail.

\subsection{LMC \hii\ Regions versus SNRs}

We obtained the scanned data to obtain integrated spectra to 
investigate the possibility that velocity broadening of emission lines may
be used to discriminate between SNRs and \hii\ regions in distant galaxies.
As seen in Table~2, the FWHM of simple, single-component Gaussians toward LMC 
\hii\ regions ranges between 1.1--1.5\,\AA\ for the \ha\ line, 1.1--1.5\,\AA\ 
for the \nii\ 6583\AA\ line, and 1.1--1.6\,\AA\ for the \sii\ 6716\,\AA\ line.
These values are close to the instrumental profile of 1.2\,\AA\ for our Goodman
spectrograph configuration.  

For the scanned data on LMC SNRs, we see that the FWHM of the emission lines
varies between 1.6--5.6\,\AA\ in \ha, 1.6--5.5\,\AA\ for the 6583\,\AA\ \nii\ 
line, and 1.8--5.0\,\AA\ for the 6716\,\AA\ \sii\ line.\footnote{Formal Gaussian fits  to the N132D spectrum give somewhat narrower values, due to ``contamination" from the strong \hii\ emission in which the SNR is embedded.  However, a much broader component for all these lines is evident in the spectrum (see Fig.~4).}  In general, the FWHMs 
of the emission-lines seen in the scanned SNR spectra are greater than the 
FWHM of those lines in LMC \hii\ regions and significantly broader than the 
instrumental profile of 1.2\,\AA. 

Even though we see some outliers in the SNR line profiles, it is clear that 
we systematically see velocity broadening in the SNRs in comparison with the 
\hii\ regions.  This is true despite the fact that we did not observe at high 
enough resolution to actually resolve the \hii\ region emission lines, which 
should be in the 10 -- 20 $\VEL$ range ($\approx$ 0.2--0.5\AA).  One expects 
bulk motions, even in older SNRs nearing the end of their radiative phase, 
of at least 50 $\VEL$.  While some of this motion will be in the plane of 
the sky, by observing the entire emission from a given object, we see 
sufficient bulk motion along the line-of-sight to broaden the lines in the 
SNRs.  Hence, observations at $\sim$50$\VEL$ resolution appear to be sufficient
to be a good diagnostic of shocks versus photoionization.  

We note in passing that  \sii:\ha\ is $\lesssim 0.4$  in the scanned spectra of both N103B (0.27) and DEML205 (0.38). Both are well-established as SNRs, and both have velocity widths that are 
%1.66 \AA\ and 4.03 \AA\ that are 
significantly broader than \hii\ regions despite the lower ratios.  While long-slit spectra may show individual filaments in these objects have higher \sii:\ha\ ratios than these global averages, the scanned spectra are more representative of the situation in more distant galaxies where long slit spectra cover much if not all of the SNR.

To assess the robustness of our conclusion that velocity broadening is a good way to separate SNRs from \hii\ regions in our sample of LMC and M83 objects, we performed an Anderson-Darling 
k-sample test on the measured FWHMs of the \hii\ regions and SNRs.  Specifically, we tested whether it was possible that the FWHMs of nebular lines of the \hii\ regions and SNRs could arise from the same distribution. In testing, we used the measured FWHMs of the scanned spectra of LMC objects and the long-slit spectra of M83 objects.  Regardless of whether we considered \ha, \nii, or \sii\ lines for the test, we found that the probability (p-value) that the samples were drawn from a common distribution was $\ll 0.001$, confirming that the quality and velocity resolution of our data is sufficient to show that the FWHMs of emission lines 
from SNRs and \hii\ regions are distinct.

\begin{comment}
\subsection{Comparison of the Velocity Broadening to \sii:\ha}

\todo[inline]{Examples in the  \sii:\ha\ $\lesssim 0.4$ regime in our sample 
consist of N103B and (0.27) and DEML205 (0.38). Can look at ratios and discuss
and can look at FWHMs and discuss.  N103B has very broad lines.  It has the
5th widest \ha\ FWHM in our sample. DEML205 has smaller velocity broadening of
1.66A still wider than all \hii\ regions. Maybe new figure with these two?}

\todo[inline]{A different way to approach this may be the following: Our paper has primarily been concerned with verifying the kinematic diagnostic.  To do that, we have concentrated on traditional optical SNRs with elevated \sii:\ha\ ratios. In the process, we have observed a couple of objects with more marginal \sii:\ha\ ratios, and the kinematic criterion still works.  This bodes well for the use of this diagnostic as we observe uncertain SNR candidates in the future.  WPB }
\end{comment}

\subsection{Integrated Line Flux Measurements and Luminosity Comparison 
with Other Galaxies}

For our spectra scanned over entire objects, we have the opportunity to obtain
{\em integrated} line fluxes, and corresponding luminosities, for SNRs in the
LMC.  After sky subtraction and extraction of 1-D spectra for each object, we
obtained the integrated flux in \ha\ (and other lines).  But since only a small
fraction of the emitting region lies within the slit at any time during the
scan, we must then divide by an ``inefficiency factor": ($slit\
area{\rm)}/{\rm(}scan\ area{\rm)} = {\rm(}slit\ width{\rm)}/{\rm(}scan\
distance{\rm)}$. For example, for N4 the 0.6\arcsec\ slit scanned at
400\,\arcsec\,hr$^{-1}$ during exposures of 1800~s (Table 1), so the fraction
of the scan covered at any instant is $0.6\arcsec/200\arcsec = 0.003$.   The
results for the 15 LMC SNRs with scanned spectra are given in Table 2.  We
measure \ha\ luminosities from $4\times 10^{35}\ {\rm to}\ 2.4\times 
10^{37}\LUM$
(assuming a distance of 50 kpc and no absorption). 

\begin{figure}
%\plotone{luminosity_comparison2.pdf}
\plotone{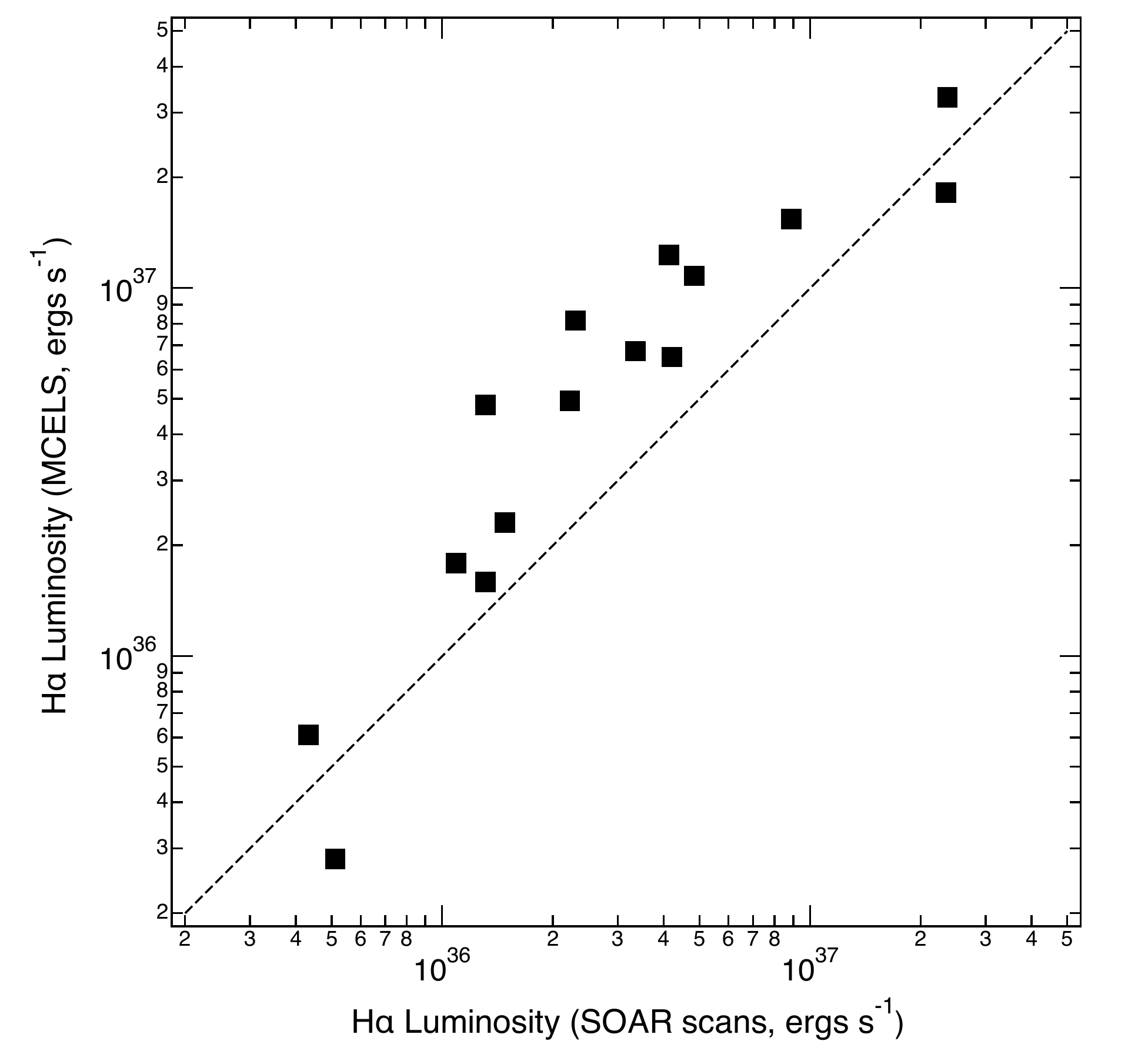}
\caption{ Comparison of the \ha\ luminosity for LMC SNRs in our sample, based on integrated flux from our scanned spectra (horizontal axis), with measurements from the MCELS continuum-subtracted \ha\ mosaic (vertical axis).  The luminosities assume a distance of 50 kpc and negligible absorption.}
\label{fig:luminosity}
\end{figure}

As a check on these values, in Figure~\ref{fig:luminosity} we compare them with
those obtained by summing over regions containing the SNRs in narrow-band \ha\
mosaic images of the LMC from MCELS \citep{smith99, paredes15}.  (As noted
above, we used these images in selecting targets for scanned spectra.)  We find
that fluxes (and luminosities, of course) are fairly well correlated in these
two independent data sets, though the MCELS image-based values are typically
larger than those from scanned spectra by about a factor of 2.  This can be
attributed to at least three factors: (1) the MCELS \ha\ filter was broad 
enough (FWHM 30\,\AA) to transmit \nii\ emission at the $\sim 30\%$ level, contaminating the \ha\ emission; 
(2) the regions measured in the MCELS images included somewhat more emission 
than what was covered in the Goodman spectrograph scans; and (3) the presence 
of non-uniform diffuse \ha\ emission toward the SNRs complicated 
sky-subtraction such that the sky-background was never completely removed.  
Each of these three effects would produce an over-estimate of the MCELS-derived \ha\ luminosities of the SNRs in comparison to the values determined 
using the scanned Goodman data.

It is interesting to compare these values with those for more distant galaxies,
where the integrated fluxes have been obtained through narrow-band imaging
and/or through long-slit spectra.\footnote{Neither of these methods are ideal:
narrow-band images through most ``\ha" filters pass some or all of the \NiiL\
line flux, while typical long slits admit only a fraction of the light from all
but the smallest objects.}  For SNRs in M33, \citet{long10} measured  \ha\
fluxes using narrow-band images from the Local Group Galaxies Survey
\citep[LGGS,][]{massey07b}, and found values up to a maximum of $2.7\times
10^{37}\LUM$.  \citet{lee14a} also used \ha\ images from the LGGS to do similar
luminosity measurements for M31, where they found SNRs (and candidates) to have
\ha\ luminosities up to $1.0 \times 10^{37} \LUM$.  

For M83, SNR line fluxes have been measured both in narrow-band images from
Magellan \citep{blair12} and in multi-slit spectra from Gemini-S
\citep{winkler17}.  As Fig. 5 of the latter paper shows, there is a tight
correlation between the two data sets; the vast majority of objects have \ha\
fluxes from imaging between 1 and 2 times those measured from spectra.  The
luminosities for M83 SNRs range up to $\sim 5\times 10^{37} \LUM$, a few times
brighter than the most luminous ones in M31 or M33.   This is probably
attributable to the fact that active star formation is taking place in M83,
with an attendant higher mean ISM density compared with the more quiescent M31
or M33 \citep{winkler17}.  

Within the context of these three much larger galaxies, we see that the
brightest SNRs in the LMC have luminosity higher than those in M31, comparable
with those in M33, and a factor of about two lower than those in M83.  We note,
however, that we do {\em not} present a complete luminosity function for LMC
SNRs, for we have sampled only about 25\% of them.  Since we intentionally
selected objects with high surface brightness, our sample should be at least
close to the bright end of the luminosity function; however, there may be other
high-luminosity objects that we did not include because they are too large to
be covered by the 3\farcm9 Goodman slit.  Nevertheless, we can conclude that
the brightest SNRs in the LMC have luminosities that at least approach those in
M83.  
% WPB: Commented out next sentence.  It doesn't really add much and will only engender a question or comment from the referee as to what it means.
%This may result from episodes of star formation in the LMC in the not-too-distant past.

\subsection{Comparison with M83 SNRs and H~II regions}

At the distance of M83 (4.6 Mpc), 1\arcsec\ corresponds with about 22 pc, so
our 0.6\arcsec\ slit width corresponds to $\sim$13 pc.  This is large enough to
engulf the majority of flux from small objects that are well-centered on the
slit, but does not get all of the flux from some objects whose diameters range
upwards of 20 pc, as is the case for a few of the objects observed.  Seeing
effects could also cause some light to be lost from the narrow slit.

Nonetheless, comparison of the kinematics for the small number of M83 SNRs in
Table 2 with the randomly sampled \hii\ regions that were observed
serendipitously along the  same slits, we see that all of the SNRs observed
show significant broadening above that seen for \hii\ regions.  For two
objects, B12-039 and B12-208, the shock broadening is subtle in the \ha\
measurements, but is more clear from the other listed lines. 
This is an indication that overlying diffuse or contaminating \ha\ emission may
be influencing the observed \ha\ despite our attempts to subtract it.  Objects
B12-023, -180, and -191 are clearly broadened in all three listed lines,
although again, one can see a systematic effect where the broadening is higher
in \nii\ and \sii\ than in \ha.  This indicates that using lines other than
\ha\ may be the most effective application of the new kinematic technique when
observing more distant galaxies, especially as the technique is applied to 
even fainter objects where correcting for \ha\ contamination may be even more 
of a problem.

\subsection{Line Widths and Shock Models}

\begin{figure}
%\plottwo{fig_ha_width_diam.pdf}{fig_ha_width_age.pdf}
\plottwo{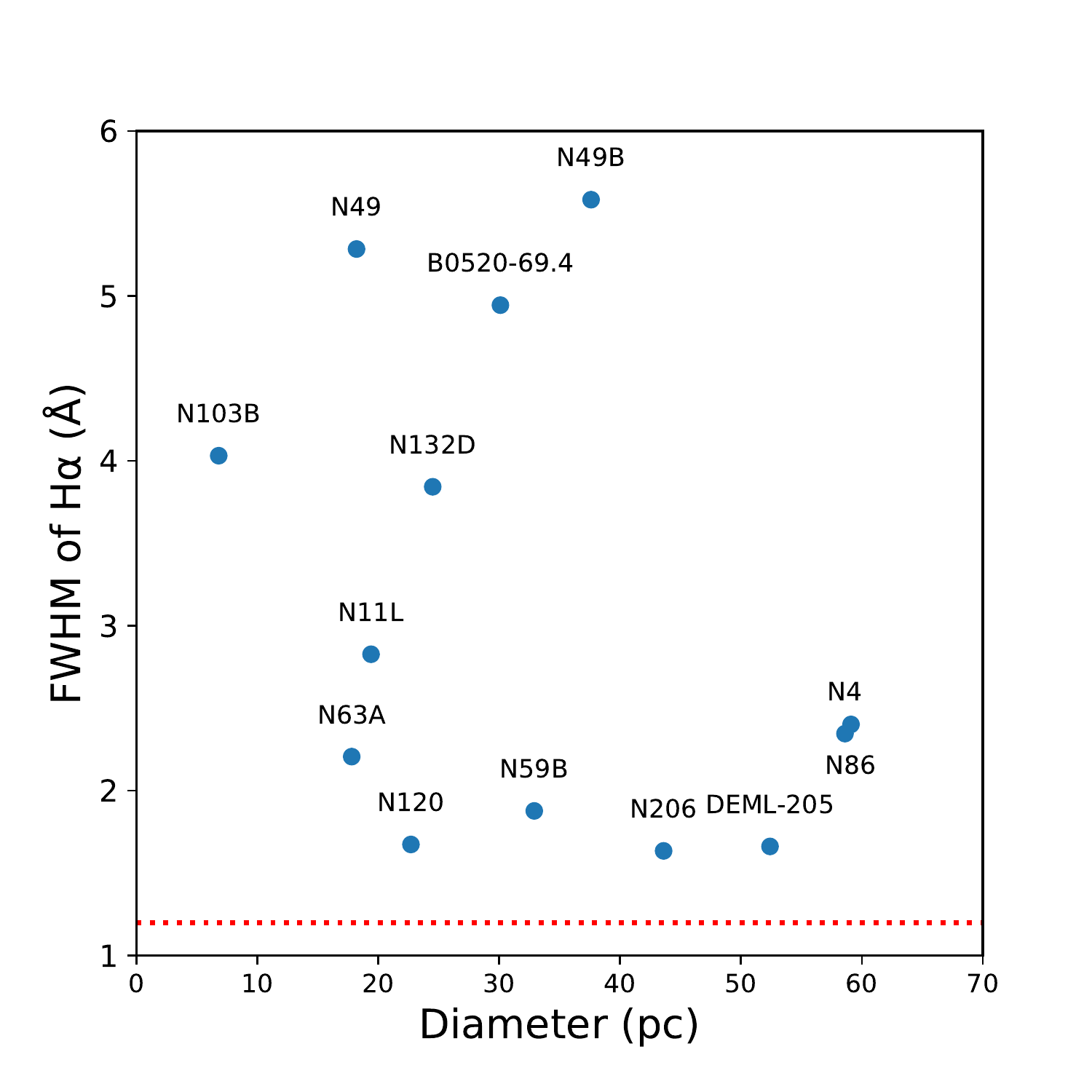}{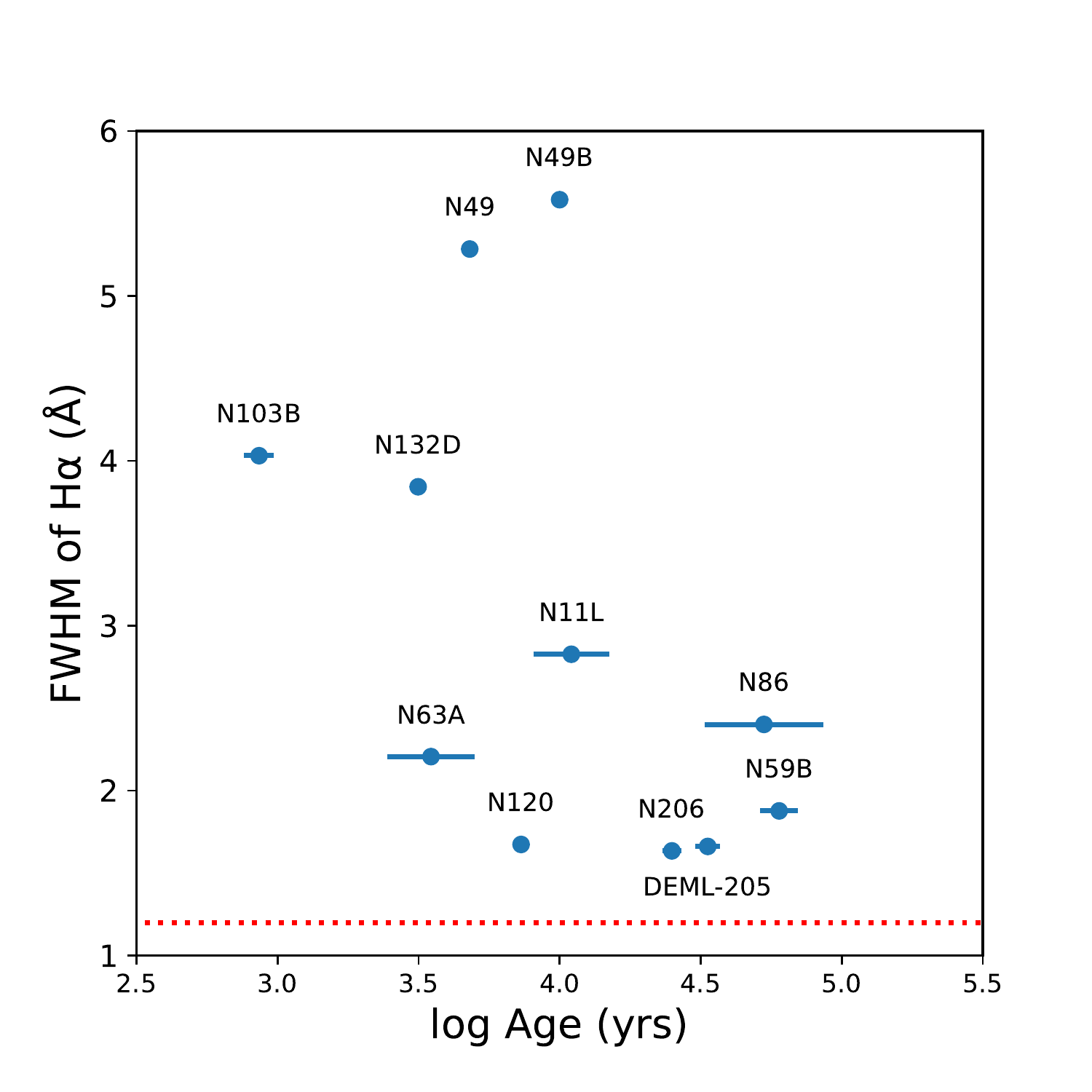}
\caption{Left:  The measured Gaussian FWHM of \ha\ as a function of SNR
diameter.  Right: The \ha\ Gaussian FWHM as a function of the age.  Data for
both x-axes, including errors on the age estimates, if available,  are as compiled by \citet{bozzetto17}. As in Fig.\ \ref{fig:widths}, the dotted line shows the instrumental resolution of 1.2 \AA. \label{fig:dia_age}
}
\end{figure}

In addition to being a discriminant between \hii\ regions and SNRs, higher
resolution observations also communicate something about the SNR itself.  The
LMC SNRs we have observed range in diameter from  approximately 6.8 pc in the
case of N103B to nearly 60 pc in the case of N86.  Analysis of the X-ray
spectrum of N103B suggests it is a relatively young, 860-year-old SNR produced
by a thermonuclear explosion \citep{hughes95}, while N86 is an old SNR,
dominated by emission from swept-up interstellar material,  with an age of
perhaps 53,000 years \citep{williams99}.  Given this, it is not surprising that
the  widths of the lines in N103B are greater than in N86.  However, as shown
in Fig.\  \ref{fig:dia_age}, any trends of line width with diameter or age are
fairly weak.  All very large diameter, very old SNRs do have fairly narrow line
widths, but objects with diameters  $\lesssim 30$ pc and ages $\lesssim$~10,000
years show a large range in observed line widths.  

On the other hand, it is important to remember that at  6500 \AA, a velocity of
$300 \kms$ corresponds to 6.5 \AA,  so the velocity widths we observe are far
less than the primary shock velocity (especially for the young objects), even
if one allows for the fact that the FWHM is less than the full widths at zero
intensity.  Qualitatively, we can understand this as follows: Consider a SNR
with a primary shock velocity of  $1000 \kms$ expanding into an ISM with a mean
density of 1 cm$^{-3}$.  The primary shock will be non-radiative, i.e.,
material behind the shock will not have had time to cool to $\sim$\,10,000\,K,
as needed to produce bright \ha\  and the forbidden lines typically observed in
optically identified SNRs.  There are SNRs where these fast shocks are
observed, the so-called Balmer-dominated SNRs \citep[see,
e.g.][]{heng10}.\footnote{N103B does have Balmer-dominated filaments
\citep{ghavamian17}, but the bulk of the radiation arises from radiative
shocks.}  In these, the only observed optical radiation   arises from H atoms undergoing ionization,
since the temperature behind the shock is 10$^7$\,K or
more.  A few such SNRs do exist in the LMC, but they are faint and are not
among the ones we have observed.  Furthermore, objects such as these are too faint to
be observed in more distant galaxies.   

In a typical SNR, the optical emission we detect arises as the primary shock
drives secondary shocks into clouds with densities of 10-100 times that of the
average ISM.   The pressure behind these secondary shocks will be in
approximate equilibrium with that of the primary shock, that is  $v_2 \sim
\sqrt{(n_1/n_2)}\:v_1$) or 100 to 300 $\kms$.  The initial  temperature behind
these shocks is still rather high, of order 10$^6$\,K, but at these
temperatures and densities the shocked gas will cool relatively quickly to the point
where it produces  significant amounts of \ha\ and forbidden-line radiation.
So except in  cases where we are viewing the ejecta directly, as  in Cas A, we
do not expect the lines to be much broader than we have observed here.

\section{Summary/Conclusions}

Large numbers of SNRs have been and are continuing to be identified in nearby
galaxies using ratios derived from narrow-band filter imaging to distinguish
SNRs from the principal source of contamination, i.e, \hii\ regions.  Because
all of the SNRs in a nearby galaxy lie at essentially the same distance and
because absorption along the line-of-sight is generally low and less variable
than in the plane of the Galaxy, such samples are important for understanding
the class properties of SNRs.  To carry out such studies accurately, however,
one must be sure that the sample is not contaminated by objects that are not
SNRs.  Here we suggest that one very clear way to eliminate imposters from a
sample is to take spectra with sufficient spectral resolution ($\sim 50 \kms$)
to separate SNRs from \hii\ regions.   

To test this suggestion, we have obtained  spectra of 15 SNRs and of 8 \hii\
regions in the LMC with the SOAR telescope with a spectral resolution of
$53\kms$. To ensure that our spectra  of the LMC objects are similar to those
typically obtained from objects in more distant galaxies, we scanned
the slit across each entire LMC object, in order to obtain ``global''  spectra.
In a few cases, we also obtained simple long-slit spectra at a single position
to compare with our scans.  We also obtained a few spectra of SNRs and \hii\
regions in M83 as a preliminary test of this technique for a more distant
galaxy.  Our conclusions are as follows: 

\begin{itemize}
\item Spectroscopy with  $\sim$50 $\VEL$ resolution is indeed an effective way to
separate SNRs from \hii\ regions, even for the largest LMC SNRs in our sample.
The typical Gaussian velocity widths of the SNRs are  $\sim$130$\VEL$
(FWHM)\footnote{This is the median value of the \ha\ widths as measured from
the Gaussian fits.}  whereas the \hii\ region lines appear near the instrumental
resolution of $\sim$50$\VEL$.  

\item The widths of lines in the ``global'' (scanned) spectra are generally
larger than the lines in corresponding simple long-slit spectra,  
because of bulk velocity shifts in the various portions of the SNR. The 
velocity widths of the lines in our spectra from H$\alpha$, \nii, and \sii\ 
are all similar
for objects with little contaminating \hii\ region emission. For the M83 SNRs,
such contamination does seem to affect  \ha\  more than the \nii\ and \sii\
lines, making these forbidden lines that are typically strong in SNRs somewhat
more effective as the kinematic diagnostic. 

\item The integrated \ha\ luminosity for LMC SNRs ranges up to at least $3 \times 10^{37}\LUM$, somewhat higher than for M31 or M33 SNRs, and approaching the values found in the starburst galaxy M83.

\item While the smaller diameter SNRs often have higher velocity widths in
their global spectra, there is considerable scatter at small diameter,
consistent with the idea that density (and possibly explosion energy) strongly
affect the appearance of SNRs at the same diameter.
\end{itemize}

In summary, we believe that kinematics provide a clean way to
discriminate between SNRs and \hii\ regions at optical wavelengths.  We plan, 
and encourage others, to pursue higher
resolution spectroscopy of the various SNR samples in nearby galaxies as a way
of resolving the uncertain identifications of objects, especially those near
the often-used \sii:\ha~=~0.4 boundary where the largest uncertainty lies.
This will result in more complete, less contaminated samples of {\em bona fide} SNRs, which are needed
for understanding the global properties of SNRs and their impact on their host
galaxies.

\acknowledgments
This work was based on observations obtained at the Southern Astrophysical
Research (SOAR) telescope, which is a joint project of the Minist\'{e}rio da
Ci\^{e}ncia, Tecnologia, Inova\c{c}\~{o}es e Comunica\c{c}\~{o}es (MCTIC) do
Brasil, the U.S. National Optical Astronomy Observatory (NOAO), the University
of North Carolina at Chapel Hill (UNC), and Michigan State University (MSU).
SDP would like to thank P. Ugarte, C. Corco, and S. Pizarro for their help at
the telescope without which these observations would not have been possible.
PFW acknowledges support from the NSF through grant AST-1714281. WPB thanks the
Center for Astrophysical Sciences at Johns Hopkins University for support.

\vspace{5mm}
\facility{SOAR (Goodman)}

%% Similar to \facility{}, there is the optional \software command to allow 
%% authors a place to specify which programs were used during the creation of 
%% the manuscript. Authors should list each code and include either a
%% citation or url to the code inside ()s when available.

%\software{astropy \citep{astropy} 
%          Cloudy \citep{2013RMxAA..49..137F}, 
 %         SExtractor \citep{1996A&AS..117..393B}

\bibliographystyle{aasjournal}

\bibliography{lmc_snrs}

\begin{thebibliography}{}
\expandafter\ifx\csname natexlab\endcsname\relax\def\natexlab#1{#1}\fi
\providecommand{\url}[1]{\href{#1}{#1}}
\providecommand{\dodoi}[1]{doi:~\href{http://doi.org/#1}{\nolinkurl{#1}}}
\providecommand{\doeprint}[1]{\href{http://ascl.net/#1}{\nolinkurl{http://ascl.net/#1}}}
\providecommand{\doarXiv}[1]{\href{https://arxiv.org/abs/#1}{\nolinkurl{https://arxiv.org/abs/#1}}}

\bibitem[{{Blair} \& {Long}(1997)}]{blair97}
{Blair}, W.~P., \& {Long}, K.~S. 1997, \apjs, 108, 261, \dodoi{10.1086/312958}

\bibitem[{{Blair} \& {Long}(2004)}]{blair04}
---. 2004, \apjs, 155, 101, \dodoi{10.1086/423958}

\bibitem[{{Blair} {et~al.}(2012){Blair}, {Winkler}, \& {Long}}]{blair12}
{Blair}, W.~P., {Winkler}, P.~F., \& {Long}, K.~S. 2012, The Astrophysical
  Journal Supplement Series, 203, 8, \dodoi{10.1088/0067-0049/203/1/8}

\bibitem[{{Bozzetto} {et~al.}(2017){Bozzetto}, {Filipovi{\'c}}, {Vukoti{\'c}},
  {Pavlovi{\'c}}, {Uro{\v s}evi{\'c}}, {Kavanagh}, {Arbutina}, {Maggi},
  {Sasaki}, {Haberl}, {Crawford}, {Roper}, {Grieve}, \& {Points}}]{bozzetto17}
{Bozzetto}, L.~M., {Filipovi{\'c}}, M.~D., {Vukoti{\'c}}, B., {et~al.} 2017,
  \apjs, 230, 2, \dodoi{10.3847/1538-4365/aa653c}

\bibitem[{{Clemens} {et~al.}(2004){Clemens}, {Crain}, \&
  {Anderson}}]{clemens04}
{Clemens}, J.~C., {Crain}, J.~A., \& {Anderson}, R. 2004, in Society of
  Photo-Optical Instrumentation Engineers (SPIE) Conference Series, Vol. 5492,
  \procspie, ed. A.~F.~M. {Moorwood} \& M.~{Iye}, 331--340

\bibitem[{{Davies} {et~al.}(1976){Davies}, {Elliott}, \& {Meaburn}}]{Davies76}
{Davies}, R.~D., {Elliott}, K.~H., \& {Meaburn}, J. 1976, \memras, 81, 89

\bibitem[{{de Vaucouleurs} {et~al.}(1983){de Vaucouleurs}, {Pence}, \&
  {Davoust}}]{deVaucouleurs83}
{de Vaucouleurs}, G., {Pence}, W.~D., \& {Davoust}, E. 1983, \apjs, 53, 17,
  \dodoi{10.1086/190881}

\bibitem[{{D'Odorico} {et~al.}(1978){D'Odorico}, {Benvenuti}, \&
  {Sabbadin}}]{dodorico78}
{D'Odorico}, S., {Benvenuti}, P., \& {Sabbadin}, F. 1978, \aap, 63, 63

\bibitem[{{Ghavamian} {et~al.}(2017){Ghavamian}, {Seitenzahl}, {Vogt},
  {Dopita}, {Terry}, {Williams}, \& {Winkler}}]{ghavamian17}
{Ghavamian}, P., {Seitenzahl}, I.~R., {Vogt}, F. P.~A., {et~al.} 2017, \apj,
  847, 122, \dodoi{10.3847/1538-4357/aa83b8}

\bibitem[{{Gordon} {et~al.}(1998){Gordon}, {Kirshner}, {Long}, {Blair},
  {Duric}, \& {Smith}}]{gordon98}
{Gordon}, S.~M., {Kirshner}, R.~P., {Long}, K.~S., {et~al.} 1998, \apjs, 117,
  89, \dodoi{10.1086/313107}

\bibitem[{{Green}(2017)}]{green17}
{Green}, D.~A. 2017, VizieR Online Data Catalog, VII/278

\bibitem[{{Haffner} {et~al.}(1999){Haffner}, {Reynolds}, \&
  {Tufte}}]{haffner99}
{Haffner}, L.~M., {Reynolds}, R.~J., \& {Tufte}, S.~L. 1999, \apj, 523, 223,
  \dodoi{10.1086/307734}

\bibitem[{{Hamuy} {et~al.}(1994){Hamuy}, {Suntzeff}, {Heathcote}, {Walker},
  {Gigoux}, \& {Phillips}}]{hamuy94}
{Hamuy}, M., {Suntzeff}, N.~B., {Heathcote}, S.~R., {et~al.} 1994, \pasp, 106,
  566

\bibitem[{{Hamuy} {et~al.}(1992){Hamuy}, {Walker}, {Suntzeff}, {Gigoux},
  {Heathcote}, \& {Phillips}}]{hamuy92}
{Hamuy}, M., {Walker}, A.~R., {Suntzeff}, N.~B., {et~al.} 1992, \pasp, 104, 533

\bibitem[{{Heng}(2010)}]{heng10}
{Heng}, K. 2010, \pasa, 27, 23, \dodoi{10.1071/AS09057}

\bibitem[{{Henize}(1956)}]{Henize56}
{Henize}, K.~G. 1956, \apjs, 2, 315, \dodoi{10.1086/190025}

\bibitem[{{Hughes} {et~al.}(1995){Hughes}, {Hayashi}, {Helfand}, {Hwang},
  {Itoh}, {Kirshner}, {Koyama}, {Markert}, {Tsunemi}, \& {Woo}}]{hughes95}
{Hughes}, J.~P., {Hayashi}, I., {Helfand}, D., {et~al.} 1995, \apjl, 444, L81,
  \dodoi{10.1086/187865}

\bibitem[{{Lee} \& {Lee}(2014{\natexlab{a}})}]{lee14a}
{Lee}, J.~H., \& {Lee}, M.~G. 2014{\natexlab{a}}, \apj, 786, 130,
  \dodoi{10.1088/0004-637X/786/2/130}

\bibitem[{{Lee} \& {Lee}(2014{\natexlab{b}})}]{lee14b}
---. 2014{\natexlab{b}}, \apj, 793, 134, \dodoi{10.1088/0004-637X/793/2/134}

\bibitem[{{Lee} {et~al.}(2015){Lee}, {Sohn}, {Lee}, {Lim}, {Jang}, {Ko}, {Koo},
  {Hwang}, {Kim}, \& {Park}}]{lee15}
{Lee}, M.~G., {Sohn}, J., {Lee}, J.~H., {et~al.} 2015, \apj, 804, 63,
  \dodoi{10.1088/0004-637X/804/1/63}

\bibitem[{{Long}(2017)}]{long17}
{Long}, K.~S. 2017, {Galactic and Extragalactic Samples of Supernova Remnants:
  How They Are Identified and What They Tell Us}, 2005

\bibitem[{{Long} {et~al.}(2018){Long}, {Blair}, {Milisavljevic}, {Raymond}, \&
  {Winkler}}]{long18}
{Long}, K.~S., {Blair}, W.~P., {Milisavljevic}, D., {Raymond}, J.~C., \&
  {Winkler}, P.~F. 2018, \apj, 855, 140, \dodoi{10.3847/1538-4357/aaac7e}

\bibitem[{{Long} {et~al.}(2019){Long}, {Winkler}, \& {Blair}}]{long19}
{Long}, K.~S., {Winkler}, P.~F., \& {Blair}, W.~P. 2019, \apj, 875, 85,
  \dodoi{10.3847/1538-4357/ab0d94}

\bibitem[{{Long} {et~al.}(2010){Long}, {Blair}, {Winkler}, {Becker}, {Gaetz},
  {Ghavamian}, {Helfand}, {Hughes}, {Kirshner}, {Kuntz}, {McNeil}, {Pannuti},
  {Plucinsky}, {Saul}, {T{\"u}llmann}, \& {Williams}}]{long10}
{Long}, K.~S., {Blair}, W.~P., {Winkler}, P.~F., {et~al.} 2010, \apjs, 187,
  495, \dodoi{10.1088/0067-0049/187/2/495}

\bibitem[{{Maddox} {et~al.}(2006){Maddox}, {Cowan}, {Kilgard}, {Lacey},
  {Prestwich}, {Stockdale}, \& {Wolfing}}]{Maddox06}
{Maddox}, L.~A., {Cowan}, J.~J., {Kilgard}, R.~E., {et~al.} 2006, \aj, 132,
  310, \dodoi{10.1086/505024}

\bibitem[{{Maggi} {et~al.}(2016){Maggi}, {Haberl}, {Kavanagh}, {Sasaki},
  {Bozzetto}, {Filipovi{\'c}}, {Vasilopoulos}, {Pietsch}, {Points}, {Chu},
  {Dickel}, {Ehle}, {Williams}, \& {Greiner}}]{maggi16}
{Maggi}, P., {Haberl}, F., {Kavanagh}, P.~J., {et~al.} 2016, \aap, 585, A162,
  \dodoi{10.1051/0004-6361/201526932}

\bibitem[{{Massey} {et~al.}(2007){Massey}, {McNeill}, {Olsen}, {Hodge},
  {Blaha}, {Jacoby}, {Smith}, \& {Strong}}]{massey07b}
{Massey}, P., {McNeill}, R.~T., {Olsen}, K.~A.~G., {et~al.} 2007, \aj, 134,
  2474, \dodoi{10.1086/523658}

\bibitem[{{Mathewson} \& {Healey}(1964)}]{mathewson64}
{Mathewson}, D.~S., \& {Healey}, J.~R. 1964, in IAU Symposium, Vol.~20, The
  Galaxy and the Magellanic Clouds, ed. F.~J. {Kerr}, 283

\bibitem[{{Matonick} \& {Fesen}(1997)}]{matonick97}
{Matonick}, D.~M., \& {Fesen}, R.~A. 1997, \apjs, 112, 49,
  \dodoi{10.1086/313034}

\bibitem[{{Paredes} {et~al.}(2015){Paredes}, {Points}, {Smith}, {Rest},
  {Damke}, {Zenteno}, \& {MCELS Team}}]{paredes15}
{Paredes}, L., {Points}, S.~D., {Smith}, R.~C., {et~al.} 2015, in Astronomical
  Society of the Pacific Conference Series, Vol. 491, Fifty Years of Wide Field
  Studies in the Southern Hemisphere: Resolved Stellar Populations of the
  Galactic Bulge and Magellanic Clouds, ed. S.~{Points} \& A.~{Kunder},
  366--369

\bibitem[{{Pietrzy{\'n}ski} {et~al.}(2013){Pietrzy{\'n}ski}, {Graczyk},
  {Gieren}, {Thompson}, {Pilecki}, {Udalski}, {Soszy{\'n}ski}, {Koz{\l}owski},
  {Konorski}, {Suchomska}, {Bono}, {Moroni}, {Villanova}, {Nardetto},
  {Bresolin}, {Kudritzki}, {Storm}, {Gallenne}, {Smolec}, {Minniti}, {Kubiak},
  {Szyma{\'n}ski}, {Poleski}, {Wyrzykowski}, {Ulaczyk}, {Pietrukowicz},
  {G{\'o}rski}, \& {Karczmarek}}]{pietrzynski13}
{Pietrzy{\'n}ski}, G., {Graczyk}, D., {Gieren}, W., {et~al.} 2013, \nat, 495,
  76, \dodoi{10.1038/nature11878}

\bibitem[{{Saha} {et~al.}(2006){Saha}, {Thim}, {Tammann}, {Reindl}, \& {Sand
  age}}]{saha06}
{Saha}, A., {Thim}, F., {Tammann}, G.~A., {Reindl}, B., \& {Sand age}, A. 2006,
  \apjs, 165, 108, \dodoi{10.1086/503800}

\bibitem[{{Smith} \& {MCELS Team}(1999)}]{smith99}
{Smith}, R.~C., \& {MCELS Team}. 1999, in IAU Symposium, Vol. 190, New Views of
  the Magellanic Clouds, ed. Y.-H. {Chu}, N.~{Suntzeff}, J.~{Hesser}, \&
  D.~{Bohlender}, 28

\bibitem[{{Soria} \& {Wu}(2003)}]{Soria03}
{Soria}, R., \& {Wu}, K. 2003, \aap, 410, 53,
  \dodoi{10.1051/0004-6361:20031074}

\bibitem[{{Westerlund} \& {Mathewson}(1966)}]{westerlund66}
{Westerlund}, B.~E., \& {Mathewson}, D.~S. 1966, \mnras, 131, 371,
  \dodoi{10.1093/mnras/131.3.371}

\bibitem[{{Williams} {et~al.}(1999){Williams}, {Chu}, {Dickel}, {Petre},
  {Smith}, \& {Tavarez}}]{williams99}
{Williams}, R.~M., {Chu}, Y.-H., {Dickel}, J.~R., {et~al.} 1999, \apjs, 123,
  467, \dodoi{10.1086/313246}

\bibitem[{{Winkler} {et~al.}(2017){Winkler}, {Blair}, \& {Long}}]{winkler17}
{Winkler}, P.~F., {Blair}, W.~P., \& {Long}, K.~S. 2017, \apj, 839, 83,
  \dodoi{10.3847/1538-4357/aa683d}

\bibitem[{{Winkler} {et~al.}(2015){Winkler}, {Smith}, {Points}, \& {MCELS
  Team}}]{winkler15}
{Winkler}, P.~F., {Smith}, R.~C., {Points}, S.~D., \& {MCELS Team}. 2015, in
  Astronomical Society of the Pacific Conference Series, Vol. 491, Fifty Years
  of Wide Field Studies in the Southern Hemisphere: Resolved Stellar
  Populations of the Galactic Bulge and Magellanic Clouds, ed. S.~{Points} \&
  A.~{Kunder}, 343

\end{thebibliography}
%\bibliography{../bibmaster.bib}

% Begin FIGURES

% Now Figure 1:

%Now Figure 2:

%Now Fig. 3

%Now Fig. 4

%Now Fig. 5

%Now Fig. 6

%Now Fig. 7

%Now Fig. 8

%Now Fig. 9

\end{document}